\documentclass[a4paper,article]{memoir}

\usepackage[]{textcomp}
\usepackage[]{microtype}
\usepackage[utf8]{inputenc}
\usepackage[T1]{fontenc}
\usepackage[]{mathtools}

\usepackage{hyperref}

\input{style.tex}

\begin{document}
  \title{Notes on axiomatising Hurkens's Paradox}\author{Arnaud Spiwack}{\maketitle}
  \begin{abstract}
    An axiomatisation of Hurkens's paradox in dependent type theory is given without assuming any impredicative feature of said type theory.
  \end{abstract}
  Hurkens's paradox~\cite{Hurkens1995} is a very economic, though rather hard to understand, paradox of the $\mbox{\textrm{U}}^-$ impredicative type theory, described in Section~\ref{latex_lib_label_2}, whose main characteristic is to feature to nested impredicative sorts. Its terseness makes it the weapon of choice to derive inconsistencies from logical principle or experimental language features of your favourite proof assistant. Or, rather, embedding $\mbox{\textrm{U}}^-$ in some way is the weapon of choice, Hurkens's paradox serves as a way to turn this into a proof of false.
  \par
  It may sound like a futile game to play: if you are the ideal mathematician you will never implement inconsistent feature in your proof assistant. Unfortunately, you are not, and deriving contradiction will happen to you from time to time. Having a tool for that may turn out to be of tremendous help. As a bonus, the inconsistency of $\mbox{\textrm{U}}^-$ can serve to derive potentially useful principles, such as the fact that if the principle of excluded middle holds in an impredicative sort, then types in that sort have the proof irrelevance property (see Section~\ref{latex_lib_label_6}).
  \par
  The downside in all that is that there does not seem to be a good way to express, within dependent type theory, the existence of an impredicative sort. Coquand~\cite{Coquand86} gave a sufficient condition, albeit much stronger, to derive contradictions in a generic way. His proof was based on Girard's~\cite{Girard1972} paradox rather than Hurkens's one (which came out ten years later). Geuvers~\cite{Geuvers2007} later gave a proof based more directly on Hurkens's one and relying on a single impredicative sort, but this proof wasn't very generic. The result was that Hurkens's proof was included \emph{twice} in the distribution of the Coq proof assistant~\cite{Coq}: Geuvers's proof, and a variant due to Hugo Herbelin to prove slightly different results.
  \par
  This situation is certainly unsatisfactory, as adapting Hurkens's proof for every little variation around the same theme is significantly more work than describing an encoding of $\mbox{\textrm{U}}^-$. It prevents good people from finding perfectly good proof of contradictions: it isn't fair to assume that everyone is an expert in Hurkens's proof.
  \par
  As it happens, however, there is a perfectly good axiomatisation of $\mbox{\textrm{U}}^-$ in your favourite dependently typed proof assistant (in actuality, a sufficient subsystem). And the corresponding proof of contradiction is, \emph{mutatis mutandis}, Geuvers's, where conversion rules are replaced by equalities.
  \chapter{Axiomatic Hurkens's paradox}\label{latex_lib_label_1}
  \par
  The trick, so to speak, of the axiomatix presentation of $\mbox{\textrm{U}}^-$ is generally attributed to Martin-Löf: a universe is given by an type {\usefont{T1}{pag}{m}{n}{\small{U}}}\symbol{58}{\usefont{T1}{pag}{b}{n}{\small{Type}}} describing the types in the universe, and an decoding function {\usefont{T1}{pag}{m}{n}{\small{El}}}\symbol{58}{\usefont{T1}{pag}{m}{n}{\small{U}}}${\rightarrow}${\usefont{T1}{pag}{b}{n}{\small{Type}}} describing, for each type in {\usefont{T1}{pag}{m}{n}{\small{U}}} the elements of that types. Sorts are to be encoded as such universes. System $\mbox{\textrm{U}}^-$ has two of these, commonly called \emph{large} and \emph{small}, together with rules to combine them. Each of these rules take the form of a product formation rule (see Barendregt's presentations of \emph{pure type systems}, formerly known as \emph{generalised type systems}~\cite{Barendregt1991}\cite[Section~5.2]{Barendregt}). Instead of the usual presentation where there is a single dependent product with a number of formation rules, we will have a distinct dependent product~--~with its own introduction rule (${\lambda}$-abstraction) and elimination rule (application)~--~for each of the formation rule. For each pair ${\lambda}$-abstraction \& application, there may be a ${\beta}$-equivalence rule, modelled as an equality; only the ${\beta}$-equivalence rules which are effectively used in the proof are axiomatised.
  \par
  \section{Axiomatic $\mbox{\textrm{U}}^-$}\label{latex_lib_label_2}
  \par
  The full axiomatic presentation appears below, in Coq syntax. It is also part of Coq's distribution and can be found, at the time these notes are being written, in the file \textsf{theories/Logic/Hurkens.v}.
  \par
  \paragraph{Large universe}
  \par
  \begin{displaymath}
    \makebox[1.\linewidth]{\makebox[0.1\linewidth]{}\parbox{0.9\linewidth}{{\usefont{T1}{pag}{b}{n}{\small{Variable}}}~{\usefont{T1}{pag}{m}{n}{\small{U1}}}~\symbol{58}~{\usefont{T1}{pag}{b}{n}{\small{Type}}}\symbol{46}\\
    {\usefont{T1}{pag}{b}{n}{\small{Variable}}}~{\usefont{T1}{pag}{m}{n}{\small{El1}}}~\symbol{58}~{\usefont{T1}{pag}{m}{n}{\small{U1}}}~${\rightarrow}$~{\usefont{T1}{pag}{b}{n}{\small{Type}}}\symbol{46}}}
  \end{displaymath}
  \par
  The large universe {\usefont{T1}{pag}{m}{n}{\small{U1}}} is closed by dependent products over types in {\usefont{T1}{pag}{m}{n}{\small{U1}}}. The definition of dependent product and ${\lambda}$-abstraction are defined using the function space of the dependent type theory. Notations are defined for dependent product, ${\lambda}$-abstraction and application. As usual, an arrow notation is used when the dependent product has a constant range.
  \par
  \begin{displaymath}
    \makebox[1.\linewidth]{\makebox[0.1\linewidth]{}\parbox{0.9\linewidth}{{\usefont{T1}{pag}{b}{n}{\small{Variable}}}~{\usefont{T1}{pag}{m}{n}{\small{Forall1}}}~\symbol{58}~{\usefont{T1}{pag}{b}{n}{\small{forall}}}~{\usefont{T1}{pag}{m}{n}{\small{u}}}\symbol{58}{\usefont{T1}{pag}{m}{n}{\small{U1}}}\symbol{44}~\symbol{40}{\usefont{T1}{pag}{m}{n}{\small{El1}}}~{\usefont{T1}{pag}{m}{n}{\small{u}}}~${\rightarrow}$~{\usefont{T1}{pag}{m}{n}{\small{U1}}}\symbol{41}~${\rightarrow}$~{\usefont{T1}{pag}{m}{n}{\small{U1}}}\symbol{46}\\
    \hphantom{ }\hphantom{ }{\usefont{T1}{pag}{b}{n}{\small{Notation}}}~\symbol{34}\symbol{39}${\forall}$$_1$\symbol{39}~{\usefont{T1}{pag}{m}{n}{\small{x}}}~\symbol{58}~{\usefont{T1}{pag}{m}{n}{\small{A}}}~\symbol{44}~{\usefont{T1}{pag}{m}{n}{\small{B}}}\symbol{34}~$\coloneqq $~\symbol{40}{\usefont{T1}{pag}{m}{n}{\small{Forall1}}}~{\usefont{T1}{pag}{m}{n}{\small{A}}}~\symbol{40}{\usefont{T1}{pag}{b}{n}{\small{fun}}}~{\usefont{T1}{pag}{m}{n}{\small{x}}}~${\Rightarrow}$~{\usefont{T1}{pag}{m}{n}{\small{B}}}\symbol{41}\symbol{41}\symbol{46}\\
    \hphantom{ }\hphantom{ }{\usefont{T1}{pag}{b}{n}{\small{Notation}}}~\symbol{34}{\usefont{T1}{pag}{m}{n}{\small{A}}}~\symbol{39}${\longrightarrow}$$_1$\symbol{39}~{\usefont{T1}{pag}{m}{n}{\small{B}}}\symbol{34}~$\coloneqq $~\symbol{40}{\usefont{T1}{pag}{m}{n}{\small{Forall1}}}~{\usefont{T1}{pag}{m}{n}{\small{A}}}~\symbol{40}{\usefont{T1}{pag}{b}{n}{\small{fun}}}~$\_$~${\Rightarrow}$~{\usefont{T1}{pag}{m}{n}{\small{B}}}\symbol{41}\symbol{41}\symbol{46}\\
    {\usefont{T1}{pag}{b}{n}{\small{Variable}}}~{\usefont{T1}{pag}{m}{n}{\small{lam1}}}~\symbol{58}~{\usefont{T1}{pag}{b}{n}{\small{forall}}}~{\usefont{T1}{pag}{m}{n}{\small{u}}}~{\usefont{T1}{pag}{m}{n}{\small{B}}}\symbol{44}~\symbol{40}{\usefont{T1}{pag}{b}{n}{\small{forall}}}~{\usefont{T1}{pag}{m}{n}{\small{x}}}\symbol{58}{\usefont{T1}{pag}{m}{n}{\small{El1}}}~{\usefont{T1}{pag}{m}{n}{\small{u}}}\symbol{44}~{\usefont{T1}{pag}{m}{n}{\small{El1}}}~\symbol{40}{\usefont{T1}{pag}{m}{n}{\small{B}}}~{\usefont{T1}{pag}{m}{n}{\small{x}}}\symbol{41}\symbol{41}~${\rightarrow}$~{\usefont{T1}{pag}{m}{n}{\small{El1}}}~\symbol{40}${\forall}$$_1$~{\usefont{T1}{pag}{m}{n}{\small{x}}}\symbol{58}{\usefont{T1}{pag}{m}{n}{\small{u}}}\symbol{44}~{\usefont{T1}{pag}{m}{n}{\small{B}}}~{\usefont{T1}{pag}{m}{n}{\small{x}}}\symbol{41}\symbol{46}\\
    \hphantom{ }\hphantom{ }{\usefont{T1}{pag}{b}{n}{\small{Notation}}}~\symbol{34}\symbol{39}${\lambda}$$_1$\symbol{39}~{\usefont{T1}{pag}{m}{n}{\small{x}}}~\symbol{44}~{\usefont{T1}{pag}{m}{n}{\small{u}}}\symbol{34}~$\coloneqq $~\symbol{40}{\usefont{T1}{pag}{m}{n}{\small{lam1}}}~$\_$~$\_$~\symbol{40}{\usefont{T1}{pag}{b}{n}{\small{fun}}}~{\usefont{T1}{pag}{m}{n}{\small{x}}}~${\Rightarrow}$~{\usefont{T1}{pag}{m}{n}{\small{u}}}\symbol{41}\symbol{41}\symbol{46}\\
    {\usefont{T1}{pag}{b}{n}{\small{Variable}}}~{\usefont{T1}{pag}{m}{n}{\small{app1}}}~\symbol{58}~{\usefont{T1}{pag}{b}{n}{\small{forall}}}~{\usefont{T1}{pag}{m}{n}{\small{u}}}~{\usefont{T1}{pag}{m}{n}{\small{B}}}~\symbol{40}{\usefont{T1}{pag}{m}{n}{\small{f}}}\symbol{58}{\usefont{T1}{pag}{m}{n}{\small{El1}}}~\symbol{40}${\forall}$$_1$~{\usefont{T1}{pag}{m}{n}{\small{x}}}\symbol{58}{\usefont{T1}{pag}{m}{n}{\small{u}}}\symbol{44}~{\usefont{T1}{pag}{m}{n}{\small{B}}}~{\usefont{T1}{pag}{m}{n}{\small{x}}}\symbol{41}\symbol{41}~\symbol{40}{\usefont{T1}{pag}{m}{n}{\small{x}}}\symbol{58}{\usefont{T1}{pag}{m}{n}{\small{El1}}}~{\usefont{T1}{pag}{m}{n}{\small{u}}}\symbol{41}\symbol{44}~{\usefont{T1}{pag}{m}{n}{\small{El1}}}~\symbol{40}{\usefont{T1}{pag}{m}{n}{\small{B}}}~{\usefont{T1}{pag}{m}{n}{\small{x}}}\symbol{41}\symbol{46}\\
    \hphantom{ }\hphantom{ }{\usefont{T1}{pag}{b}{n}{\small{Notation}}}~\symbol{34}{\usefont{T1}{pag}{m}{n}{\small{f}}}~\symbol{39}${\cdot}$$_1$\symbol{39}~{\usefont{T1}{pag}{m}{n}{\small{x}}}\symbol{34}~$\coloneqq $~\symbol{40}{\usefont{T1}{pag}{m}{n}{\small{app1}}}~$\_$~$\_$~{\usefont{T1}{pag}{m}{n}{\small{f}}}~{\usefont{T1}{pag}{m}{n}{\small{x}}}\symbol{41}\symbol{46}\\
    {\usefont{T1}{pag}{b}{n}{\small{Variable}}}~{\usefont{T1}{pag}{m}{n}{\small{beta1}}}~\symbol{58}~{\usefont{T1}{pag}{b}{n}{\small{forall}}}~{\usefont{T1}{pag}{m}{n}{\small{u}}}~{\usefont{T1}{pag}{m}{n}{\small{B}}}~\symbol{40}{\usefont{T1}{pag}{m}{n}{\small{f}}}\symbol{58}{\usefont{T1}{pag}{b}{n}{\small{forall}}}~{\usefont{T1}{pag}{m}{n}{\small{x}}}\symbol{58}{\usefont{T1}{pag}{m}{n}{\small{El1}}}~{\usefont{T1}{pag}{m}{n}{\small{u}}}\symbol{44}~{\usefont{T1}{pag}{m}{n}{\small{El1}}}~\symbol{40}{\usefont{T1}{pag}{m}{n}{\small{B}}}~{\usefont{T1}{pag}{m}{n}{\small{x}}}\symbol{41}\symbol{41}~{\usefont{T1}{pag}{m}{n}{\small{x}}}\symbol{44}\\
    \hphantom{ }\hphantom{ }\hphantom{ }\hphantom{ }\hphantom{ }\hphantom{ }\hphantom{ }\hphantom{ }\hphantom{ }\hphantom{ }\hphantom{ }\hphantom{ }\hphantom{ }\hphantom{ }\hphantom{ }\hphantom{ }\hphantom{ }\hphantom{ }\hphantom{ }\symbol{40}${\lambda}$$_1$~{\usefont{T1}{pag}{m}{n}{\small{y}}}\symbol{44}~{\usefont{T1}{pag}{m}{n}{\small{f}}}~{\usefont{T1}{pag}{m}{n}{\small{y}}}\symbol{41}~${\cdot}$$_1$~{\usefont{T1}{pag}{m}{n}{\small{x}}}~\symbol{61}~{\usefont{T1}{pag}{m}{n}{\small{f}}}~{\usefont{T1}{pag}{m}{n}{\small{x}}}\symbol{46}}}
  \end{displaymath}
  \par
  The large universe {\usefont{T1}{pag}{m}{n}{\small{U1}}} is made impredicative by a dependent product with large domain. The standard presentation would use a sort {\usefont{T1}{pag}{m}{n}{\small{U2}}}, of which {\usefont{T1}{pag}{m}{n}{\small{U1}}} is a member; the dependent product would then have, as a domain, some {\usefont{T1}{pag}{m}{n}{\small{T}}}\symbol{58}{\usefont{T1}{pag}{m}{n}{\small{U2}}}. This would be unnecessary complexity as {\usefont{T1}{pag}{m}{n}{\small{U2}}} is so restricted that the only interesting type in it would be {\usefont{T1}{pag}{m}{n}{\small{U1}}}. So, instead, we simply restrict the domain of the product to be {\usefont{T1}{pag}{m}{n}{\small{U1}}}.
  \par
  \begin{displaymath}
    \makebox[1.\linewidth]{\makebox[0.1\linewidth]{}\parbox{0.9\linewidth}{{\usefont{T1}{pag}{b}{n}{\small{Variable}}}~{\usefont{T1}{pag}{m}{n}{\small{ForallU1}}}~\symbol{58}~\symbol{40}{\usefont{T1}{pag}{m}{n}{\small{U1}}}${\rightarrow}${\usefont{T1}{pag}{m}{n}{\small{U1}}}\symbol{41}~${\rightarrow}$~{\usefont{T1}{pag}{m}{n}{\small{U1}}}\symbol{46}\\
    \hphantom{ }\hphantom{ }{\usefont{T1}{pag}{b}{n}{\small{Notation}}}~\symbol{34}\symbol{39}${\forall}$$_2$\symbol{39}~{\usefont{T1}{pag}{m}{n}{\small{A}}}~\symbol{44}~{\usefont{T1}{pag}{m}{n}{\small{F}}}\symbol{34}~$\coloneqq $~\symbol{40}{\usefont{T1}{pag}{m}{n}{\small{ForallU1}}}~\symbol{40}{\usefont{T1}{pag}{b}{n}{\small{fun}}}~{\usefont{T1}{pag}{m}{n}{\small{A}}}~${\Rightarrow}$~{\usefont{T1}{pag}{m}{n}{\small{F}}}\symbol{41}\symbol{41}\symbol{46}\\
    {\usefont{T1}{pag}{b}{n}{\small{Variable}}}~{\usefont{T1}{pag}{m}{n}{\small{lamU1}}}~\symbol{58}~{\usefont{T1}{pag}{b}{n}{\small{forall}}}~{\usefont{T1}{pag}{m}{n}{\small{F}}}\symbol{44}~\symbol{40}{\usefont{T1}{pag}{b}{n}{\small{forall}}}~{\usefont{T1}{pag}{m}{n}{\small{A}}}\symbol{58}{\usefont{T1}{pag}{m}{n}{\small{U1}}}\symbol{44}~{\usefont{T1}{pag}{m}{n}{\small{El1}}}~\symbol{40}{\usefont{T1}{pag}{m}{n}{\small{F}}}~{\usefont{T1}{pag}{m}{n}{\small{A}}}\symbol{41}\symbol{41}~${\rightarrow}$~{\usefont{T1}{pag}{m}{n}{\small{El1}}}~\symbol{40}${\forall}$$_2$~{\usefont{T1}{pag}{m}{n}{\small{A}}}\symbol{44}~{\usefont{T1}{pag}{m}{n}{\small{F}}}~{\usefont{T1}{pag}{m}{n}{\small{A}}}\symbol{41}\symbol{46}\\
    \hphantom{ }\hphantom{ }{\usefont{T1}{pag}{b}{n}{\small{Notation}}}~\symbol{34}\symbol{39}${\lambda}$$_2$\symbol{39}~{\usefont{T1}{pag}{m}{n}{\small{x}}}~\symbol{44}~{\usefont{T1}{pag}{m}{n}{\small{u}}}\symbol{34}~$\coloneqq $~\symbol{40}{\usefont{T1}{pag}{m}{n}{\small{lamU1}}}~$\_$~\symbol{40}{\usefont{T1}{pag}{b}{n}{\small{fun}}}~{\usefont{T1}{pag}{m}{n}{\small{x}}}~${\Rightarrow}$~{\usefont{T1}{pag}{m}{n}{\small{u}}}\symbol{41}\symbol{41}\symbol{46}\\
    {\usefont{T1}{pag}{b}{n}{\small{Variable}}}~{\usefont{T1}{pag}{m}{n}{\small{appU1}}}~\symbol{58}~{\usefont{T1}{pag}{b}{n}{\small{forall}}}~{\usefont{T1}{pag}{m}{n}{\small{F}}}~\symbol{40}{\usefont{T1}{pag}{m}{n}{\small{f}}}\symbol{58}{\usefont{T1}{pag}{m}{n}{\small{El1}}}\symbol{40}${\forall}$$_2$~{\usefont{T1}{pag}{m}{n}{\small{A}}}\symbol{44}{\usefont{T1}{pag}{m}{n}{\small{F}}}~{\usefont{T1}{pag}{m}{n}{\small{A}}}\symbol{41}\symbol{41}~\symbol{40}{\usefont{T1}{pag}{m}{n}{\small{A}}}\symbol{58}{\usefont{T1}{pag}{m}{n}{\small{U1}}}\symbol{41}\symbol{44}~{\usefont{T1}{pag}{m}{n}{\small{El1}}}~\symbol{40}{\usefont{T1}{pag}{m}{n}{\small{F}}}~{\usefont{T1}{pag}{m}{n}{\small{A}}}\symbol{41}\symbol{46}\\
    \hphantom{ }\hphantom{ }{\usefont{T1}{pag}{b}{n}{\small{Notation}}}~\symbol{34}{\usefont{T1}{pag}{m}{n}{\small{f}}}~\symbol{39}${\cdot}$$_1$\symbol{39}~\symbol{91}~{\usefont{T1}{pag}{m}{n}{\small{A}}}~\symbol{93}\symbol{34}~$\coloneqq $~\symbol{40}{\usefont{T1}{pag}{m}{n}{\small{appU1}}}~$\_$~{\usefont{T1}{pag}{m}{n}{\small{f}}}~{\usefont{T1}{pag}{m}{n}{\small{A}}}\symbol{41}\symbol{46}\\
    {\usefont{T1}{pag}{b}{n}{\small{Variable}}}~{\usefont{T1}{pag}{m}{n}{\small{betaU1}}}~\symbol{58}~{\usefont{T1}{pag}{b}{n}{\small{forall}}}~{\usefont{T1}{pag}{m}{n}{\small{F}}}~\symbol{40}{\usefont{T1}{pag}{m}{n}{\small{f}}}\symbol{58}{\usefont{T1}{pag}{b}{n}{\small{forall}}}~{\usefont{T1}{pag}{m}{n}{\small{A}}}\symbol{58}{\usefont{T1}{pag}{m}{n}{\small{U1}}}\symbol{44}~{\usefont{T1}{pag}{m}{n}{\small{El1}}}~\symbol{40}{\usefont{T1}{pag}{m}{n}{\small{F}}}~{\usefont{T1}{pag}{m}{n}{\small{A}}}\symbol{41}\symbol{41}~{\usefont{T1}{pag}{m}{n}{\small{A}}}\symbol{44}\\
    \hphantom{ }\hphantom{ }\hphantom{ }\hphantom{ }\hphantom{ }\hphantom{ }\hphantom{ }\hphantom{ }\hphantom{ }\hphantom{ }\hphantom{ }\hphantom{ }\hphantom{ }\hphantom{ }\hphantom{ }\hphantom{ }\hphantom{ }\hphantom{ }\hphantom{ }\hphantom{ }\symbol{40}${\lambda}$$_2$~{\usefont{T1}{pag}{m}{n}{\small{x}}}\symbol{44}~{\usefont{T1}{pag}{m}{n}{\small{f}}}~{\usefont{T1}{pag}{m}{n}{\small{x}}}\symbol{41}~${\cdot}$$_1$~\symbol{91}~{\usefont{T1}{pag}{m}{n}{\small{A}}}~\symbol{93}~\symbol{61}~{\usefont{T1}{pag}{m}{n}{\small{f}}}~{\usefont{T1}{pag}{m}{n}{\small{A}}}\symbol{46}}}
  \end{displaymath}
  \par
  \paragraph{Small universe}
  \par
  The small universe {\usefont{T1}{pag}{m}{n}{\small{U0}}} is an element of the larger one. Therefore we need an {\usefont{T1}{pag}{m}{n}{\small{u0}}}\symbol{58}{\usefont{T1}{pag}{m}{n}{\small{U1}}} and {\usefont{T1}{pag}{m}{n}{\small{U0}}} is taken to be {\usefont{T1}{pag}{m}{n}{\small{El1}}}~{\usefont{T1}{pag}{m}{n}{\small{u0}}} rather than a variable.
  \par
  \begin{displaymath}
    \makebox[1.\linewidth]{\makebox[0.1\linewidth]{}\parbox{0.9\linewidth}{{\usefont{T1}{pag}{b}{n}{\small{Variable}}}~{\usefont{T1}{pag}{m}{n}{\small{u0}}}~\symbol{58}~{\usefont{T1}{pag}{m}{n}{\small{U1}}}\symbol{46}\\
    {\usefont{T1}{pag}{b}{n}{\small{Notation}}}~{\usefont{T1}{pag}{m}{n}{\small{U0}}}~$\coloneqq $~\symbol{40}{\usefont{T1}{pag}{m}{n}{\small{El1}}}~{\usefont{T1}{pag}{m}{n}{\small{u0}}}\symbol{41}\symbol{46}\\
    {\usefont{T1}{pag}{b}{n}{\small{Variable}}}~{\usefont{T1}{pag}{m}{n}{\small{El0}}}~\symbol{58}~{\usefont{T1}{pag}{m}{n}{\small{U0}}}~${\rightarrow}$~{\usefont{T1}{pag}{b}{n}{\small{Type}}}\symbol{46}}}
  \end{displaymath}
  \par
  The small universe {\usefont{T1}{pag}{m}{n}{\small{U0}}} is closed by dependent products in {\usefont{T1}{pag}{m}{n}{\small{U0}}}. The definitions are symmetric to the corresponding ones of {\usefont{T1}{pag}{m}{n}{\small{U1}}}. Notice, however, the lack of ${\beta}$-rule, which is unnecessary to derive a contradiction.
  \par
  \begin{displaymath}
    \makebox[1.\linewidth]{\makebox[0.1\linewidth]{}\parbox{0.9\linewidth}{{\usefont{T1}{pag}{b}{n}{\small{Variable}}}~{\usefont{T1}{pag}{m}{n}{\small{Forall0}}}~\symbol{58}~{\usefont{T1}{pag}{b}{n}{\small{forall}}}~{\usefont{T1}{pag}{m}{n}{\small{u}}}\symbol{58}{\usefont{T1}{pag}{m}{n}{\small{U0}}}\symbol{44}~\symbol{40}{\usefont{T1}{pag}{m}{n}{\small{El0}}}~{\usefont{T1}{pag}{m}{n}{\small{u}}}~${\rightarrow}$~{\usefont{T1}{pag}{m}{n}{\small{U0}}}\symbol{41}~${\rightarrow}$~{\usefont{T1}{pag}{m}{n}{\small{U0}}}\symbol{46}\\
    \hphantom{ }\hphantom{ }{\usefont{T1}{pag}{b}{n}{\small{Notation}}}~\symbol{34}\symbol{39}${\forall}$$_0$\symbol{39}~{\usefont{T1}{pag}{m}{n}{\small{x}}}~\symbol{58}~{\usefont{T1}{pag}{m}{n}{\small{A}}}~\symbol{44}~{\usefont{T1}{pag}{m}{n}{\small{B}}}\symbol{34}~$\coloneqq $~\symbol{40}{\usefont{T1}{pag}{m}{n}{\small{Forall0}}}~{\usefont{T1}{pag}{m}{n}{\small{A}}}~\symbol{40}{\usefont{T1}{pag}{b}{n}{\small{fun}}}~{\usefont{T1}{pag}{m}{n}{\small{x}}}~${\Rightarrow}$~{\usefont{T1}{pag}{m}{n}{\small{B}}}\symbol{41}\symbol{41}\symbol{46}\\
    \hphantom{ }\hphantom{ }{\usefont{T1}{pag}{b}{n}{\small{Notation}}}~\symbol{34}{\usefont{T1}{pag}{m}{n}{\small{A}}}~\symbol{39}${\longrightarrow}$$_0$\symbol{39}~{\usefont{T1}{pag}{m}{n}{\small{B}}}\symbol{34}~$\coloneqq $~\symbol{40}{\usefont{T1}{pag}{m}{n}{\small{Forall0}}}~{\usefont{T1}{pag}{m}{n}{\small{A}}}~\symbol{40}{\usefont{T1}{pag}{b}{n}{\small{fun}}}~$\_$~${\Rightarrow}$~{\usefont{T1}{pag}{m}{n}{\small{B}}}\symbol{41}\symbol{41}\symbol{46}\\
    {\usefont{T1}{pag}{b}{n}{\small{Variable}}}~{\usefont{T1}{pag}{m}{n}{\small{lam0}}}~\symbol{58}~{\usefont{T1}{pag}{b}{n}{\small{forall}}}~{\usefont{T1}{pag}{m}{n}{\small{u}}}~{\usefont{T1}{pag}{m}{n}{\small{B}}}\symbol{44}~\symbol{40}{\usefont{T1}{pag}{b}{n}{\small{forall}}}~{\usefont{T1}{pag}{m}{n}{\small{x}}}\symbol{58}{\usefont{T1}{pag}{m}{n}{\small{El0}}}~{\usefont{T1}{pag}{m}{n}{\small{u}}}\symbol{44}~{\usefont{T1}{pag}{m}{n}{\small{El0}}}~\symbol{40}{\usefont{T1}{pag}{m}{n}{\small{B}}}~{\usefont{T1}{pag}{m}{n}{\small{x}}}\symbol{41}\symbol{41}~${\rightarrow}$~{\usefont{T1}{pag}{m}{n}{\small{El0}}}~\symbol{40}${\forall}$$_0$~{\usefont{T1}{pag}{m}{n}{\small{x}}}\symbol{58}{\usefont{T1}{pag}{m}{n}{\small{u}}}\symbol{44}~{\usefont{T1}{pag}{m}{n}{\small{B}}}~{\usefont{T1}{pag}{m}{n}{\small{x}}}\symbol{41}\symbol{46}\\
    \hphantom{ }\hphantom{ }{\usefont{T1}{pag}{b}{n}{\small{Notation}}}~\symbol{34}\symbol{39}${\lambda}$$_0$\symbol{39}~{\usefont{T1}{pag}{m}{n}{\small{x}}}~\symbol{44}~{\usefont{T1}{pag}{m}{n}{\small{u}}}\symbol{34}~$\coloneqq $~\symbol{40}{\usefont{T1}{pag}{m}{n}{\small{lam0}}}~$\_$~$\_$~\symbol{40}{\usefont{T1}{pag}{b}{n}{\small{fun}}}~{\usefont{T1}{pag}{m}{n}{\small{x}}}~${\Rightarrow}$~{\usefont{T1}{pag}{m}{n}{\small{u}}}\symbol{41}\symbol{41}\symbol{46}\\
    {\usefont{T1}{pag}{b}{n}{\small{Variable}}}~{\usefont{T1}{pag}{m}{n}{\small{app0}}}~\symbol{58}~{\usefont{T1}{pag}{b}{n}{\small{forall}}}~{\usefont{T1}{pag}{m}{n}{\small{u}}}~{\usefont{T1}{pag}{m}{n}{\small{B}}}~\symbol{40}{\usefont{T1}{pag}{m}{n}{\small{f}}}\symbol{58}{\usefont{T1}{pag}{m}{n}{\small{El0}}}~\symbol{40}${\forall}$$_0$~{\usefont{T1}{pag}{m}{n}{\small{x}}}\symbol{58}{\usefont{T1}{pag}{m}{n}{\small{u}}}\symbol{44}~{\usefont{T1}{pag}{m}{n}{\small{B}}}~{\usefont{T1}{pag}{m}{n}{\small{x}}}\symbol{41}\symbol{41}~\symbol{40}{\usefont{T1}{pag}{m}{n}{\small{x}}}\symbol{58}{\usefont{T1}{pag}{m}{n}{\small{El0}}}~{\usefont{T1}{pag}{m}{n}{\small{u}}}\symbol{41}\symbol{44}~{\usefont{T1}{pag}{m}{n}{\small{El0}}}~\symbol{40}{\usefont{T1}{pag}{m}{n}{\small{B}}}~{\usefont{T1}{pag}{m}{n}{\small{x}}}\symbol{41}\symbol{46}\\
    \hphantom{ }\hphantom{ }{\usefont{T1}{pag}{b}{n}{\small{Notation}}}~\symbol{34}{\usefont{T1}{pag}{m}{n}{\small{f}}}~\symbol{39}${\cdot}$$_0$\symbol{39}~{\usefont{T1}{pag}{m}{n}{\small{x}}}\symbol{34}~$\coloneqq $~\symbol{40}{\usefont{T1}{pag}{m}{n}{\small{app0}}}~$\_$~$\_$~{\usefont{T1}{pag}{m}{n}{\small{f}}}~{\usefont{T1}{pag}{m}{n}{\small{x}}}\symbol{41}\symbol{46}}}
  \end{displaymath}
  \par
  The small universe {\usefont{T1}{pag}{m}{n}{\small{U0}}} is made impredicative by a dependent whose range is in {\usefont{T1}{pag}{m}{n}{\small{U1}}}. Contrary to the impredicative product of {\usefont{T1}{pag}{m}{n}{\small{U1}}}, the range cannot be restricted to be only {\usefont{T1}{pag}{m}{n}{\small{U0}}}. Here again, the ${\beta}$-rule is not needed.
  \par
  \begin{displaymath}
    \makebox[1.\linewidth]{\makebox[0.1\linewidth]{}\parbox{0.9\linewidth}{{\usefont{T1}{pag}{b}{n}{\small{Variable}}}~{\usefont{T1}{pag}{m}{n}{\small{ForallU0}}}~\symbol{58}~{\usefont{T1}{pag}{b}{n}{\small{forall}}}~{\usefont{T1}{pag}{m}{n}{\small{u}}}\symbol{58}{\usefont{T1}{pag}{m}{n}{\small{U1}}}\symbol{44}~\symbol{40}{\usefont{T1}{pag}{m}{n}{\small{El1}}}~{\usefont{T1}{pag}{m}{n}{\small{u}}}${\rightarrow}${\usefont{T1}{pag}{m}{n}{\small{U0}}}\symbol{41}~${\rightarrow}$~{\usefont{T1}{pag}{m}{n}{\small{U0}}}\symbol{46}\\
    \hphantom{ }\hphantom{ }{\usefont{T1}{pag}{b}{n}{\small{Notation}}}~\symbol{34}\symbol{39}${\forall}$$_0$$^1$\symbol{39}~{\usefont{T1}{pag}{m}{n}{\small{A}}}~\symbol{58}~{\usefont{T1}{pag}{m}{n}{\small{U}}}~\symbol{44}~{\usefont{T1}{pag}{m}{n}{\small{F}}}\symbol{34}~$\coloneqq $~\symbol{40}{\usefont{T1}{pag}{m}{n}{\small{ForallU0}}}~{\usefont{T1}{pag}{m}{n}{\small{U}}}~\symbol{40}{\usefont{T1}{pag}{b}{n}{\small{fun}}}~{\usefont{T1}{pag}{m}{n}{\small{A}}}~${\Rightarrow}$~{\usefont{T1}{pag}{m}{n}{\small{F}}}\symbol{41}\symbol{41}\symbol{46}\\
    {\usefont{T1}{pag}{b}{n}{\small{Variable}}}~{\usefont{T1}{pag}{m}{n}{\small{lamU0}}}~\symbol{58}~{\usefont{T1}{pag}{b}{n}{\small{forall}}}~{\usefont{T1}{pag}{m}{n}{\small{U}}}~{\usefont{T1}{pag}{m}{n}{\small{F}}}\symbol{44}~\symbol{40}{\usefont{T1}{pag}{b}{n}{\small{forall}}}~{\usefont{T1}{pag}{m}{n}{\small{A}}}\symbol{58}{\usefont{T1}{pag}{m}{n}{\small{El1}}}~{\usefont{T1}{pag}{m}{n}{\small{U}}}\symbol{44}~{\usefont{T1}{pag}{m}{n}{\small{El0}}}~\symbol{40}{\usefont{T1}{pag}{m}{n}{\small{F}}}~{\usefont{T1}{pag}{m}{n}{\small{A}}}\symbol{41}\symbol{41}~${\rightarrow}$~{\usefont{T1}{pag}{m}{n}{\small{El0}}}~\symbol{40}${\forall}$$_0$$^1$~{\usefont{T1}{pag}{m}{n}{\small{A}}}\symbol{58}{\usefont{T1}{pag}{m}{n}{\small{U}}}\symbol{44}~{\usefont{T1}{pag}{m}{n}{\small{F}}}~{\usefont{T1}{pag}{m}{n}{\small{A}}}\symbol{41}\symbol{46}\\
    \hphantom{ }\hphantom{ }{\usefont{T1}{pag}{b}{n}{\small{Notation}}}~\symbol{34}\symbol{39}${\lambda}$$_0$$^1$\symbol{39}~{\usefont{T1}{pag}{m}{n}{\small{x}}}~\symbol{44}~{\usefont{T1}{pag}{m}{n}{\small{u}}}\symbol{34}~$\coloneqq $~\symbol{40}{\usefont{T1}{pag}{m}{n}{\small{lamU0}}}~$\_$~$\_$~\symbol{40}{\usefont{T1}{pag}{b}{n}{\small{fun}}}~{\usefont{T1}{pag}{m}{n}{\small{x}}}~${\Rightarrow}$~{\usefont{T1}{pag}{m}{n}{\small{u}}}\symbol{41}\symbol{41}\symbol{46}\\
    {\usefont{T1}{pag}{b}{n}{\small{Variable}}}~{\usefont{T1}{pag}{m}{n}{\small{appU0}}}~\symbol{58}~{\usefont{T1}{pag}{b}{n}{\small{forall}}}~{\usefont{T1}{pag}{m}{n}{\small{U}}}~{\usefont{T1}{pag}{m}{n}{\small{F}}}~\symbol{40}{\usefont{T1}{pag}{m}{n}{\small{f}}}\symbol{58}{\usefont{T1}{pag}{m}{n}{\small{El0}}}\symbol{40}${\forall}$$_0$$^1$~{\usefont{T1}{pag}{m}{n}{\small{A}}}\symbol{58}{\usefont{T1}{pag}{m}{n}{\small{U}}}\symbol{44}{\usefont{T1}{pag}{m}{n}{\small{F}}}~{\usefont{T1}{pag}{m}{n}{\small{A}}}\symbol{41}\symbol{41}~\symbol{40}{\usefont{T1}{pag}{m}{n}{\small{A}}}\symbol{58}{\usefont{T1}{pag}{m}{n}{\small{El1}}}~{\usefont{T1}{pag}{m}{n}{\small{U}}}\symbol{41}\symbol{44}~{\usefont{T1}{pag}{m}{n}{\small{El0}}}~\symbol{40}{\usefont{T1}{pag}{m}{n}{\small{F}}}~{\usefont{T1}{pag}{m}{n}{\small{A}}}\symbol{41}\symbol{46}\\
    \hphantom{ }\hphantom{ }{\usefont{T1}{pag}{b}{n}{\small{Notation}}}~\symbol{34}{\usefont{T1}{pag}{m}{n}{\small{f}}}~\symbol{39}${\cdot}$$_0$\symbol{39}~\symbol{91}~{\usefont{T1}{pag}{m}{n}{\small{A}}}~\symbol{93}\symbol{34}~$\coloneqq $~\symbol{40}{\usefont{T1}{pag}{m}{n}{\small{appU0}}}~$\_$~$\_$~{\usefont{T1}{pag}{m}{n}{\small{f}}}~{\usefont{T1}{pag}{m}{n}{\small{A}}}\symbol{41}\symbol{46}}}
  \end{displaymath}
  \par
  \section{Proof of contradiction}
  \par
  From there, we can proceed to use Hurkens's argument to derive a contradiction. Let's be precise: we shall prove that every type in {\usefont{T1}{pag}{m}{n}{\small{U0}}} is inhabited. It will only be an actual contradiction if {\usefont{T1}{pag}{m}{n}{\small{U0}}} contains the empty type. For this purpose, let's assume a type in {\usefont{T1}{pag}{m}{n}{\small{U0}}}, we will then prove it is inhabited.
  \par
  \begin{displaymath}
    \makebox[1.\linewidth]{\makebox[0.1\linewidth]{}\parbox{0.9\linewidth}{{\usefont{T1}{pag}{b}{n}{\small{Variable}}}~{\usefont{T1}{pag}{m}{n}{\small{F}}}\symbol{58}{\usefont{T1}{pag}{m}{n}{\small{U0}}}\symbol{46}}}
  \end{displaymath}
  \par
  The proof will require simplifying ${\beta}$-redexes. We provide tactics to that effect.
  \par
  \begin{displaymath}
    \makebox[1.\linewidth]{\makebox[0.1\linewidth]{}\parbox{0.9\linewidth}{{\usefont{T1}{pag}{b}{n}{\small{Ltac}}}~{\usefont{T1}{pag}{m}{n}{\small{simplify}}}~$\coloneqq $\\
    \hphantom{ }\hphantom{ }\symbol{40}{\usefont{T1}{pag}{m}{n}{\small{{\itshape repeat}}}}~{\usefont{T1}{pag}{m}{n}{\small{{\itshape rewrite}}}}~\symbol{63}{\usefont{T1}{pag}{m}{n}{\small{beta1}}}\symbol{44}~\symbol{63}{\usefont{T1}{pag}{m}{n}{\small{betaU1}}}\symbol{41}\symbol{59}\\
    \hphantom{ }\hphantom{ }{\usefont{T1}{pag}{m}{n}{\small{{\itshape lazy}}}}~{\usefont{T1}{pag}{m}{n}{\small{beta}}}\symbol{46}\\[0.5\baselineskip]
    {\usefont{T1}{pag}{b}{n}{\small{Ltac}}}~{\usefont{T1}{pag}{m}{n}{\small{simplify\_in}}}~{\usefont{T1}{pag}{m}{n}{\small{h}}}~$\coloneqq $\\
    \hphantom{ }\hphantom{ }\symbol{40}{\usefont{T1}{pag}{m}{n}{\small{{\itshape repeat}}}}~{\usefont{T1}{pag}{m}{n}{\small{{\itshape rewrite}}}}~\symbol{63}{\usefont{T1}{pag}{m}{n}{\small{beta1}}}\symbol{44}~\symbol{63}{\usefont{T1}{pag}{m}{n}{\small{betaU1}}}~{\usefont{T1}{pag}{b}{n}{\small{in}}}~{\usefont{T1}{pag}{m}{n}{\small{h}}}\symbol{41}\symbol{59}\\
    \hphantom{ }\hphantom{ }{\usefont{T1}{pag}{m}{n}{\small{{\itshape lazy}}}}~{\usefont{T1}{pag}{m}{n}{\small{beta}}}~{\usefont{T1}{pag}{b}{n}{\small{in}}}~{\usefont{T1}{pag}{m}{n}{\small{h}}}\symbol{46}}}
  \end{displaymath}
  \par
  These tactics are rather brute-force, in that they will ${\beta}$-reduce as much as possible without any particular strategy. On the other hand, they, crucially, don't unfold Coq definitions so that we can give them hints by manually unfolding the appropriate terms to be simplified. Allowing the simplification tactics to unfold Coq definitions turns out to be intractable.
  \par
  It is traditional to regard {\usefont{T1}{pag}{m}{n}{\small{U1}}} as the type of datatypes and {\usefont{T1}{pag}{m}{n}{\small{U0}}} as the type of proposition. This view is justified by the fact that {\usefont{T1}{pag}{m}{n}{\small{U0}}} is not equipped with ${\beta}$-conversion rules. In the proof, following Geuvers~\cite{Geuvers2007}, data is explicitly given, while propositions are proved with tactics. Here are the data definitions (I'm playing a bit loose here, since I consider propositions to be data, they are according to the above definition at least):
  \par
  \begin{displaymath}
    \makebox[1.\linewidth]{\makebox[0.1\linewidth]{}\parbox{0.9\linewidth}{{\usefont{T1}{pag}{b}{n}{\small{Definition}}}~{\usefont{T1}{pag}{m}{n}{\small{V}}}~\symbol{58}~{\usefont{T1}{pag}{m}{n}{\small{U1}}}~$\coloneqq $~${\forall}$$_2$~{\usefont{T1}{pag}{m}{n}{\small{A}}}\symbol{44}~\symbol{40}\symbol{40}{\usefont{T1}{pag}{m}{n}{\small{A}}}~${\longrightarrow}$$_1$~{\usefont{T1}{pag}{m}{n}{\small{u0}}}\symbol{41}~${\longrightarrow}$$_1$~{\usefont{T1}{pag}{m}{n}{\small{A}}}~${\longrightarrow}$$_1$~{\usefont{T1}{pag}{m}{n}{\small{u0}}}\symbol{41}~${\longrightarrow}$$_1$~{\usefont{T1}{pag}{m}{n}{\small{A}}}~${\longrightarrow}$$_1$~{\usefont{T1}{pag}{m}{n}{\small{u0}}}\symbol{46}\\
    {\usefont{T1}{pag}{b}{n}{\small{Definition}}}~{\usefont{T1}{pag}{m}{n}{\small{U}}}~\symbol{58}~{\usefont{T1}{pag}{m}{n}{\small{U1}}}~$\coloneqq $~{\usefont{T1}{pag}{m}{n}{\small{V}}}~${\longrightarrow}$$_1$~{\usefont{T1}{pag}{m}{n}{\small{u0}}}\symbol{46}\\[0.5\baselineskip]
    {\usefont{T1}{pag}{b}{n}{\small{Definition}}}~{\usefont{T1}{pag}{m}{n}{\small{sb}}}~\symbol{40}{\usefont{T1}{pag}{m}{n}{\small{z}}}\symbol{58}{\usefont{T1}{pag}{m}{n}{\small{El1}}}~{\usefont{T1}{pag}{m}{n}{\small{V}}}\symbol{41}~\symbol{58}~{\usefont{T1}{pag}{m}{n}{\small{El1}}}~{\usefont{T1}{pag}{m}{n}{\small{V}}}~$\coloneqq $~${\lambda}$$_2$~{\usefont{T1}{pag}{m}{n}{\small{A}}}\symbol{44}~${\lambda}$$_1$~{\usefont{T1}{pag}{m}{n}{\small{r}}}\symbol{44}~${\lambda}$$_1$~{\usefont{T1}{pag}{m}{n}{\small{a}}}\symbol{44}~{\usefont{T1}{pag}{m}{n}{\small{r}}}~${\cdot}$$_1$~\symbol{40}{\usefont{T1}{pag}{m}{n}{\small{z}}}${\cdot}$$_1$\symbol{91}{\usefont{T1}{pag}{m}{n}{\small{A}}}\symbol{93}${\cdot}$$_1${\usefont{T1}{pag}{m}{n}{\small{r}}}\symbol{41}~${\cdot}$$_1$~{\usefont{T1}{pag}{m}{n}{\small{a}}}\symbol{46}\\[0.5\baselineskip]
    {\usefont{T1}{pag}{b}{n}{\small{Definition}}}~{\usefont{T1}{pag}{m}{n}{\small{le}}}~\symbol{40}{\usefont{T1}{pag}{m}{n}{\small{i}}}\symbol{58}{\usefont{T1}{pag}{m}{n}{\small{El1}}}~\symbol{40}{\usefont{T1}{pag}{m}{n}{\small{U}}}${\longrightarrow}$$_1${\usefont{T1}{pag}{m}{n}{\small{u0}}}\symbol{41}\symbol{41}~\symbol{40}{\usefont{T1}{pag}{m}{n}{\small{x}}}\symbol{58}{\usefont{T1}{pag}{m}{n}{\small{El1}}}~{\usefont{T1}{pag}{m}{n}{\small{U}}}\symbol{41}~\symbol{58}~{\usefont{T1}{pag}{m}{n}{\small{U0}}}~$\coloneqq $\\
    \hphantom{ }\hphantom{ }{\usefont{T1}{pag}{m}{n}{\small{x}}}~${\cdot}$$_1$~\symbol{40}${\lambda}$$_2$~{\usefont{T1}{pag}{m}{n}{\small{A}}}\symbol{44}~${\lambda}$$_1$~{\usefont{T1}{pag}{m}{n}{\small{r}}}\symbol{44}~${\lambda}$$_1$~{\usefont{T1}{pag}{m}{n}{\small{a}}}\symbol{44}~{\usefont{T1}{pag}{m}{n}{\small{i}}}~${\cdot}$$_1$~\symbol{40}${\lambda}$$_1$~{\usefont{T1}{pag}{m}{n}{\small{v}}}\symbol{44}~\symbol{40}{\usefont{T1}{pag}{m}{n}{\small{sb}}}~{\usefont{T1}{pag}{m}{n}{\small{v}}}\symbol{41}~${\cdot}$$_1$~\symbol{91}{\usefont{T1}{pag}{m}{n}{\small{A}}}\symbol{93}~${\cdot}$$_1$~{\usefont{T1}{pag}{m}{n}{\small{r}}}~${\cdot}$$_1$~{\usefont{T1}{pag}{m}{n}{\small{a}}}\symbol{41}\symbol{41}\symbol{46}\\
    {\usefont{T1}{pag}{b}{n}{\small{Definition}}}~{\usefont{T1}{pag}{m}{n}{\small{le}}}\symbol{39}~\symbol{58}~{\usefont{T1}{pag}{m}{n}{\small{El1}}}~\symbol{40}\symbol{40}{\usefont{T1}{pag}{m}{n}{\small{U}}}${\longrightarrow}$$_1${\usefont{T1}{pag}{m}{n}{\small{u0}}}\symbol{41}~${\longrightarrow}$$_1$~{\usefont{T1}{pag}{m}{n}{\small{U}}}~${\longrightarrow}$$_1$~{\usefont{T1}{pag}{m}{n}{\small{u0}}}\symbol{41}~$\coloneqq $~${\lambda}$$_1$~{\usefont{T1}{pag}{m}{n}{\small{i}}}\symbol{44}~${\lambda}$$_1$~{\usefont{T1}{pag}{m}{n}{\small{x}}}\symbol{44}~{\usefont{T1}{pag}{m}{n}{\small{le}}}~{\usefont{T1}{pag}{m}{n}{\small{i}}}~{\usefont{T1}{pag}{m}{n}{\small{x}}}\symbol{46}\\
    {\usefont{T1}{pag}{b}{n}{\small{Definition}}}~{\usefont{T1}{pag}{m}{n}{\small{induct}}}~\symbol{40}{\usefont{T1}{pag}{m}{n}{\small{i}}}\symbol{58}{\usefont{T1}{pag}{m}{n}{\small{El1}}}~\symbol{40}{\usefont{T1}{pag}{m}{n}{\small{U}}}${\longrightarrow}$$_1${\usefont{T1}{pag}{m}{n}{\small{u0}}}\symbol{41}\symbol{41}~\symbol{58}~{\usefont{T1}{pag}{m}{n}{\small{U0}}}~$\coloneqq $\\
    \hphantom{ }\hphantom{ }${\forall}$$_0$$^1$~{\usefont{T1}{pag}{m}{n}{\small{x}}}\symbol{58}{\usefont{T1}{pag}{m}{n}{\small{U}}}\symbol{44}~{\usefont{T1}{pag}{m}{n}{\small{le}}}~{\usefont{T1}{pag}{m}{n}{\small{i}}}~{\usefont{T1}{pag}{m}{n}{\small{x}}}~${\longrightarrow}$$_0$~{\usefont{T1}{pag}{m}{n}{\small{i}}}~${\cdot}$$_1$~{\usefont{T1}{pag}{m}{n}{\small{x}}}\symbol{46}\\[0.5\baselineskip]
    {\usefont{T1}{pag}{b}{n}{\small{Definition}}}~{\usefont{T1}{pag}{m}{n}{\small{WF}}}~\symbol{58}~{\usefont{T1}{pag}{m}{n}{\small{El1}}}~{\usefont{T1}{pag}{m}{n}{\small{U}}}~$\coloneqq $~${\lambda}$$_1$~{\usefont{T1}{pag}{m}{n}{\small{z}}}\symbol{44}~\symbol{40}{\usefont{T1}{pag}{m}{n}{\small{induct}}}~\symbol{40}{\usefont{T1}{pag}{m}{n}{\small{z}}}${\cdot}$$_1$\symbol{91}{\usefont{T1}{pag}{m}{n}{\small{U}}}\symbol{93}~${\cdot}$$_1$~{\usefont{T1}{pag}{m}{n}{\small{le}}}\symbol{39}\symbol{41}\symbol{41}\symbol{46}\\
    {\usefont{T1}{pag}{b}{n}{\small{Definition}}}~{\usefont{T1}{pag}{m}{n}{\small{I}}}~\symbol{40}{\usefont{T1}{pag}{m}{n}{\small{x}}}\symbol{58}{\usefont{T1}{pag}{m}{n}{\small{El1}}}~{\usefont{T1}{pag}{m}{n}{\small{U}}}\symbol{41}~\symbol{58}~{\usefont{T1}{pag}{m}{n}{\small{U0}}}~$\coloneqq $\\
    \hphantom{ }\hphantom{ }\symbol{40}${\forall}$$_0$$^1$~{\usefont{T1}{pag}{m}{n}{\small{i}}}\symbol{58}{\usefont{T1}{pag}{m}{n}{\small{U}}}${\longrightarrow}$$_1${\usefont{T1}{pag}{m}{n}{\small{u0}}}\symbol{44}~{\usefont{T1}{pag}{m}{n}{\small{le}}}~{\usefont{T1}{pag}{m}{n}{\small{i}}}~{\usefont{T1}{pag}{m}{n}{\small{x}}}~${\longrightarrow}$$_0$~{\usefont{T1}{pag}{m}{n}{\small{i}}}~${\cdot}$$_1$~\symbol{40}${\lambda}$$_1$~{\usefont{T1}{pag}{m}{n}{\small{v}}}\symbol{44}~\symbol{40}{\usefont{T1}{pag}{m}{n}{\small{sb}}}~{\usefont{T1}{pag}{m}{n}{\small{v}}}\symbol{41}~${\cdot}$$_1$~\symbol{91}{\usefont{T1}{pag}{m}{n}{\small{U}}}\symbol{93}~${\cdot}$$_1$~{\usefont{T1}{pag}{m}{n}{\small{le}}}\symbol{39}~${\cdot}$$_1$~{\usefont{T1}{pag}{m}{n}{\small{x}}}\symbol{41}\symbol{41}~${\longrightarrow}$$_0$~{\usefont{T1}{pag}{m}{n}{\small{F}}}\\
    \symbol{46}}}
  \end{displaymath}
  \par
  The proofs follow Geuvers~\cite{Geuvers2007} as well. The main difference is that we must explicitly call to {\usefont{T1}{pag}{m}{n}{\small{simplify}}} where conversion was used implicitly and that standard Coq tactics calls to the {\usefont{T1}{pag}{m}{n}{\small{{\itshape intro}}}} and {\usefont{T1}{pag}{m}{n}{\small{{\itshape apply}}}} tactics are generally replaced by tactics of the form {\usefont{T1}{pag}{m}{n}{\small{{\itshape refine}}}}~\symbol{40}${\lambda}$$_0$~{\usefont{T1}{pag}{m}{n}{\small{x}}}\symbol{44}~$\_$\symbol{41} and {\usefont{T1}{pag}{m}{n}{\small{{\itshape refine}}}}~\symbol{40}{\usefont{T1}{pag}{m}{n}{\small{h}}}${\cdot}$$_0$$\_$\symbol{41} respectively.
  \par
  \begin{displaymath}
    \makebox[1.\linewidth]{\makebox[0.1\linewidth]{}\parbox{0.9\linewidth}{{\usefont{T1}{pag}{b}{n}{\small{Lemma}}}~{\usefont{T1}{pag}{m}{n}{\small{Omega}}}~\symbol{58}~{\usefont{T1}{pag}{m}{n}{\small{El0}}}~\symbol{40}${\forall}$$_0$$^1$~{\usefont{T1}{pag}{m}{n}{\small{i}}}\symbol{58}{\usefont{T1}{pag}{m}{n}{\small{U}}}${\longrightarrow}$$_1${\usefont{T1}{pag}{m}{n}{\small{u0}}}\symbol{44}~{\usefont{T1}{pag}{m}{n}{\small{induct}}}~{\usefont{T1}{pag}{m}{n}{\small{i}}}~${\longrightarrow}$$_0$~{\usefont{T1}{pag}{m}{n}{\small{i}}}~${\cdot}$$_1$~{\usefont{T1}{pag}{m}{n}{\small{WF}}}\symbol{41}\symbol{46}\\
    {\usefont{T1}{pag}{b}{n}{\small{Proof}}}\symbol{46}\\
    \hphantom{ }\hphantom{ }{\usefont{T1}{pag}{m}{n}{\small{{\itshape refine}}}}~\symbol{40}${\lambda}$$_0$$^1$~{\usefont{T1}{pag}{m}{n}{\small{i}}}\symbol{44}~${\lambda}$$_0$~{\usefont{T1}{pag}{m}{n}{\small{y}}}\symbol{44}~$\_$\symbol{41}\symbol{46}\\
    \hphantom{ }\hphantom{ }{\usefont{T1}{pag}{m}{n}{\small{{\itshape refine}}}}~\symbol{40}{\usefont{T1}{pag}{m}{n}{\small{y}}}${\cdot}$$_0$\symbol{91}$\_$\symbol{93}${\cdot}$$_0$$\_$\symbol{41}\symbol{46}\\
    \hphantom{ }\hphantom{ }{\usefont{T1}{pag}{m}{n}{\small{{\itshape unfold}}}}~{\usefont{T1}{pag}{m}{n}{\small{le}}}\symbol{44}{\usefont{T1}{pag}{m}{n}{\small{WF}}}\symbol{44}{\usefont{T1}{pag}{m}{n}{\small{induct}}}\symbol{46}~{\usefont{T1}{pag}{m}{n}{\small{simplify}}}\symbol{46}\\
    \hphantom{ }\hphantom{ }{\usefont{T1}{pag}{m}{n}{\small{{\itshape refine}}}}~\symbol{40}${\lambda}$$_0$$^1$~{\usefont{T1}{pag}{m}{n}{\small{x}}}\symbol{44}~${\lambda}$$_0$~{\usefont{T1}{pag}{m}{n}{\small{h0}}}\symbol{44}~$\_$\symbol{41}\symbol{46}~{\usefont{T1}{pag}{m}{n}{\small{simplify}}}\symbol{46}\\
    \hphantom{ }\hphantom{ }{\usefont{T1}{pag}{m}{n}{\small{{\itshape refine}}}}~\symbol{40}{\usefont{T1}{pag}{m}{n}{\small{y}}}${\cdot}$$_0$\symbol{91}$\_$\symbol{93}${\cdot}$$_0$$\_$\symbol{41}\symbol{46}\\
    \hphantom{ }\hphantom{ }{\usefont{T1}{pag}{m}{n}{\small{{\itshape unfold}}}}~{\usefont{T1}{pag}{m}{n}{\small{le}}}\symbol{46}~{\usefont{T1}{pag}{m}{n}{\small{simplify}}}\symbol{46}~{\usefont{T1}{pag}{m}{n}{\small{{\itshape unfold}}}}~{\usefont{T1}{pag}{m}{n}{\small{sb}}}~{\usefont{T1}{pag}{m}{n}{\small{{\itshape at}}}}~1\symbol{46}~{\usefont{T1}{pag}{m}{n}{\small{simplify}}}\symbol{46}~{\usefont{T1}{pag}{m}{n}{\small{{\itshape unfold}}}}~{\usefont{T1}{pag}{m}{n}{\small{le}}}\symbol{39}~{\usefont{T1}{pag}{m}{n}{\small{{\itshape at}}}}~1\symbol{46}~{\usefont{T1}{pag}{m}{n}{\small{simplify}}}\symbol{46}\\
    \hphantom{ }\hphantom{ }{\usefont{T1}{pag}{m}{n}{\small{{\itshape exact}}}}~{\usefont{T1}{pag}{m}{n}{\small{h0}}}\symbol{46}\\
    {\usefont{T1}{pag}{b}{n}{\small{Qed}}}\symbol{46}}}
  \end{displaymath}
  \par
  \begin{displaymath}
    \makebox[1.\linewidth]{\makebox[0.1\linewidth]{}\parbox{0.9\linewidth}{{\usefont{T1}{pag}{b}{n}{\small{Lemma}}}~{\usefont{T1}{pag}{m}{n}{\small{lemma1}}}~\symbol{58}~{\usefont{T1}{pag}{m}{n}{\small{El0}}}~\symbol{40}{\usefont{T1}{pag}{m}{n}{\small{induct}}}~\symbol{40}${\lambda}$$_1$~{\usefont{T1}{pag}{m}{n}{\small{u}}}\symbol{44}~{\usefont{T1}{pag}{m}{n}{\small{I}}}~{\usefont{T1}{pag}{m}{n}{\small{u}}}\symbol{41}\symbol{41}\symbol{46}\\
    {\usefont{T1}{pag}{b}{n}{\small{Proof}}}\symbol{46}\\
    \hphantom{ }\hphantom{ }{\usefont{T1}{pag}{m}{n}{\small{{\itshape unfold}}}}~{\usefont{T1}{pag}{m}{n}{\small{induct}}}\symbol{46}\\
    \hphantom{ }\hphantom{ }{\usefont{T1}{pag}{m}{n}{\small{{\itshape refine}}}}~\symbol{40}${\lambda}$$_0$$^1$~{\usefont{T1}{pag}{m}{n}{\small{x}}}\symbol{44}~${\lambda}$$_0$~{\usefont{T1}{pag}{m}{n}{\small{p}}}\symbol{44}~$\_$\symbol{41}\symbol{46}~{\usefont{T1}{pag}{m}{n}{\small{simplify}}}\symbol{46}\\
    \hphantom{ }\hphantom{ }{\usefont{T1}{pag}{m}{n}{\small{{\itshape refine}}}}~\symbol{40}${\lambda}$$_0$~{\usefont{T1}{pag}{m}{n}{\small{q}}}\symbol{44}$\_$\symbol{41}\symbol{46}\\
    \hphantom{ }\hphantom{ }{\usefont{T1}{pag}{m}{n}{\small{{\itshape assert}}}}~\symbol{40}{\usefont{T1}{pag}{m}{n}{\small{El0}}}~\symbol{40}{\usefont{T1}{pag}{m}{n}{\small{I}}}~\symbol{40}${\lambda}$$_1$~{\usefont{T1}{pag}{m}{n}{\small{v}}}\symbol{44}~\symbol{40}{\usefont{T1}{pag}{m}{n}{\small{sb}}}~{\usefont{T1}{pag}{m}{n}{\small{v}}}\symbol{41}${\cdot}$$_1$\symbol{91}{\usefont{T1}{pag}{m}{n}{\small{U}}}\symbol{93}${\cdot}$$_1${\usefont{T1}{pag}{m}{n}{\small{le}}}\symbol{39}${\cdot}$$_1${\usefont{T1}{pag}{m}{n}{\small{x}}}\symbol{41}\symbol{41}\symbol{41}~{\usefont{T1}{pag}{m}{n}{\small{{\itshape as}}}}~{\usefont{T1}{pag}{m}{n}{\small{h}}}\symbol{46}\\
    \hphantom{ }\hphantom{ }\{~{\usefont{T1}{pag}{m}{n}{\small{{\itshape generalize}}}}~\symbol{40}{\usefont{T1}{pag}{m}{n}{\small{q}}}${\cdot}$$_0$\symbol{91}${\lambda}$$_1$~{\usefont{T1}{pag}{m}{n}{\small{u}}}\symbol{44}~{\usefont{T1}{pag}{m}{n}{\small{I}}}~{\usefont{T1}{pag}{m}{n}{\small{u}}}\symbol{93}${\cdot}$$_0${\usefont{T1}{pag}{m}{n}{\small{p}}}\symbol{41}\symbol{46}~{\usefont{T1}{pag}{m}{n}{\small{simplify}}}\symbol{46}\\
    \hphantom{ }\hphantom{ }\hphantom{ }\hphantom{ }{\usefont{T1}{pag}{m}{n}{\small{{\itshape intros}}}}~{\usefont{T1}{pag}{m}{n}{\small{q}}}\symbol{39}\symbol{46}~{\usefont{T1}{pag}{m}{n}{\small{{\itshape exact}}}}~{\usefont{T1}{pag}{m}{n}{\small{q}}}\symbol{39}\symbol{46}~\}\\
    \hphantom{ }\hphantom{ }{\usefont{T1}{pag}{m}{n}{\small{{\itshape refine}}}}~\symbol{40}{\usefont{T1}{pag}{m}{n}{\small{h}}}${\cdot}$$_0$$\_$\symbol{41}\symbol{46}\\
    \hphantom{ }\hphantom{ }{\usefont{T1}{pag}{m}{n}{\small{{\itshape refine}}}}~\symbol{40}${\lambda}$$_0$$^1$~{\usefont{T1}{pag}{m}{n}{\small{i}}}\symbol{44}$\_$\symbol{41}\symbol{46}\\
    \hphantom{ }\hphantom{ }{\usefont{T1}{pag}{m}{n}{\small{{\itshape refine}}}}~\symbol{40}${\lambda}$$_0$~{\usefont{T1}{pag}{m}{n}{\small{h}}}\symbol{39}\symbol{44}~$\_$\symbol{41}\symbol{46}\\
    \hphantom{ }\hphantom{ }{\usefont{T1}{pag}{m}{n}{\small{{\itshape generalize}}}}~\symbol{40}{\usefont{T1}{pag}{m}{n}{\small{q}}}${\cdot}$$_0$\symbol{91}${\lambda}$$_1$~{\usefont{T1}{pag}{m}{n}{\small{y}}}\symbol{44}~{\usefont{T1}{pag}{m}{n}{\small{i}}}~${\cdot}$$_1$~\symbol{40}${\lambda}$$_1$~{\usefont{T1}{pag}{m}{n}{\small{v}}}\symbol{44}~\symbol{40}{\usefont{T1}{pag}{m}{n}{\small{sb}}}~{\usefont{T1}{pag}{m}{n}{\small{v}}}\symbol{41}${\cdot}$$_1$\symbol{91}{\usefont{T1}{pag}{m}{n}{\small{U}}}\symbol{93}~${\cdot}$$_1$~{\usefont{T1}{pag}{m}{n}{\small{le}}}\symbol{39}~${\cdot}$$_1$~{\usefont{T1}{pag}{m}{n}{\small{y}}}\symbol{41}\symbol{93}\symbol{41}\symbol{46}~{\usefont{T1}{pag}{m}{n}{\small{simplify}}}\symbol{46}\\
    \hphantom{ }\hphantom{ }{\usefont{T1}{pag}{m}{n}{\small{{\itshape intros}}}}~{\usefont{T1}{pag}{m}{n}{\small{q}}}\symbol{39}\symbol{46}\\
    \hphantom{ }\hphantom{ }{\usefont{T1}{pag}{m}{n}{\small{{\itshape refine}}}}~\symbol{40}{\usefont{T1}{pag}{m}{n}{\small{q}}}\symbol{39}${\cdot}$$_0$$\_$\symbol{41}\symbol{46}~{\usefont{T1}{pag}{m}{n}{\small{{\itshape clear}}}}~{\usefont{T1}{pag}{m}{n}{\small{q}}}\symbol{39}\symbol{46}\\
    \hphantom{ }\hphantom{ }{\usefont{T1}{pag}{m}{n}{\small{{\itshape unfold}}}}~{\usefont{T1}{pag}{m}{n}{\small{le}}}~{\usefont{T1}{pag}{m}{n}{\small{{\itshape at}}}}~1~{\usefont{T1}{pag}{b}{n}{\small{in}}}~{\usefont{T1}{pag}{m}{n}{\small{h}}}\symbol{39}\symbol{46}~{\usefont{T1}{pag}{m}{n}{\small{simplify\_in}}}~{\usefont{T1}{pag}{m}{n}{\small{h}}}\symbol{39}\symbol{46}\\
    \hphantom{ }\hphantom{ }{\usefont{T1}{pag}{m}{n}{\small{{\itshape unfold}}}}~{\usefont{T1}{pag}{m}{n}{\small{sb}}}~{\usefont{T1}{pag}{m}{n}{\small{{\itshape at}}}}~1~{\usefont{T1}{pag}{b}{n}{\small{in}}}~{\usefont{T1}{pag}{m}{n}{\small{h}}}\symbol{39}\symbol{46}~{\usefont{T1}{pag}{m}{n}{\small{simplify\_in}}}~{\usefont{T1}{pag}{m}{n}{\small{h}}}\symbol{39}\symbol{46}\\
    \hphantom{ }\hphantom{ }{\usefont{T1}{pag}{m}{n}{\small{{\itshape unfold}}}}~{\usefont{T1}{pag}{m}{n}{\small{le}}}\symbol{39}~{\usefont{T1}{pag}{m}{n}{\small{{\itshape at}}}}~1~{\usefont{T1}{pag}{b}{n}{\small{in}}}~{\usefont{T1}{pag}{m}{n}{\small{h}}}\symbol{39}\symbol{46}~{\usefont{T1}{pag}{m}{n}{\small{simplify\_in}}}~{\usefont{T1}{pag}{m}{n}{\small{h}}}\symbol{39}\symbol{46}\\
    \hphantom{ }\hphantom{ }{\usefont{T1}{pag}{m}{n}{\small{{\itshape exact}}}}~{\usefont{T1}{pag}{m}{n}{\small{h}}}\symbol{39}\symbol{46}\\
    {\usefont{T1}{pag}{b}{n}{\small{Qed}}}\symbol{46}}}
  \end{displaymath}
  \par
  \begin{displaymath}
    \makebox[1.\linewidth]{\makebox[0.1\linewidth]{}\parbox{0.9\linewidth}{{\usefont{T1}{pag}{b}{n}{\small{Lemma}}}~{\usefont{T1}{pag}{m}{n}{\small{lemma2}}}~\symbol{58}~{\usefont{T1}{pag}{m}{n}{\small{El0}}}~\symbol{40}\symbol{40}${\forall}$$_0$$^1${\usefont{T1}{pag}{m}{n}{\small{i}}}\symbol{58}{\usefont{T1}{pag}{m}{n}{\small{U}}}${\longrightarrow}$$_1${\usefont{T1}{pag}{m}{n}{\small{u0}}}\symbol{44}~{\usefont{T1}{pag}{m}{n}{\small{induct}}}~{\usefont{T1}{pag}{m}{n}{\small{i}}}~${\longrightarrow}$$_0$~{\usefont{T1}{pag}{m}{n}{\small{i}}}${\cdot}$$_1${\usefont{T1}{pag}{m}{n}{\small{WF}}}\symbol{41}~${\longrightarrow}$$_0$~{\usefont{T1}{pag}{m}{n}{\small{F}}}\symbol{41}\symbol{46}\\
    {\usefont{T1}{pag}{b}{n}{\small{Proof}}}\symbol{46}\\
    \hphantom{ }\hphantom{ }{\usefont{T1}{pag}{m}{n}{\small{{\itshape refine}}}}~\symbol{40}${\lambda}$$_0$~{\usefont{T1}{pag}{m}{n}{\small{x}}}\symbol{44}~$\_$\symbol{41}\symbol{46}\\
    \hphantom{ }\hphantom{ }{\usefont{T1}{pag}{m}{n}{\small{{\itshape assert}}}}~\symbol{40}{\usefont{T1}{pag}{m}{n}{\small{El0}}}~\symbol{40}{\usefont{T1}{pag}{m}{n}{\small{I}}}~{\usefont{T1}{pag}{m}{n}{\small{WF}}}\symbol{41}\symbol{41}~{\usefont{T1}{pag}{m}{n}{\small{{\itshape as}}}}~{\usefont{T1}{pag}{m}{n}{\small{h}}}\symbol{46}\\
    \hphantom{ }\hphantom{ }\{~{\usefont{T1}{pag}{m}{n}{\small{{\itshape generalize}}}}~\symbol{40}{\usefont{T1}{pag}{m}{n}{\small{x}}}${\cdot}$$_0$\symbol{91}${\lambda}$$_1$~{\usefont{T1}{pag}{m}{n}{\small{u}}}\symbol{44}~{\usefont{T1}{pag}{m}{n}{\small{I}}}~{\usefont{T1}{pag}{m}{n}{\small{u}}}\symbol{93}${\cdot}$$_0${\usefont{T1}{pag}{m}{n}{\small{lemma1}}}\symbol{41}\symbol{46}~{\usefont{T1}{pag}{m}{n}{\small{simplify}}}\symbol{46}\\
    \hphantom{ }\hphantom{ }\hphantom{ }\hphantom{ }{\usefont{T1}{pag}{m}{n}{\small{{\itshape intros}}}}~{\usefont{T1}{pag}{m}{n}{\small{q}}}\symbol{46}\\
    \hphantom{ }\hphantom{ }\hphantom{ }\hphantom{ }{\usefont{T1}{pag}{m}{n}{\small{{\itshape exact}}}}~{\usefont{T1}{pag}{m}{n}{\small{q}}}\symbol{46}~\}\\
    \hphantom{ }\hphantom{ }{\usefont{T1}{pag}{m}{n}{\small{{\itshape refine}}}}~\symbol{40}{\usefont{T1}{pag}{m}{n}{\small{h}}}${\cdot}$$_0$$\_$\symbol{41}\symbol{46}~{\usefont{T1}{pag}{m}{n}{\small{{\itshape clear}}}}~{\usefont{T1}{pag}{m}{n}{\small{h}}}\symbol{46}\\
    \hphantom{ }\hphantom{ }{\usefont{T1}{pag}{m}{n}{\small{{\itshape refine}}}}~\symbol{40}${\lambda}$$_0$$^1$~{\usefont{T1}{pag}{m}{n}{\small{i}}}\symbol{44}~${\lambda}$$_0$~{\usefont{T1}{pag}{m}{n}{\small{h0}}}\symbol{44}~$\_$\symbol{41}\symbol{46}\\
    \hphantom{ }\hphantom{ }{\usefont{T1}{pag}{m}{n}{\small{{\itshape generalize}}}}~\symbol{40}{\usefont{T1}{pag}{m}{n}{\small{x}}}${\cdot}$$_0$\symbol{91}${\lambda}$$_1$~{\usefont{T1}{pag}{m}{n}{\small{y}}}\symbol{44}~{\usefont{T1}{pag}{m}{n}{\small{i}}}${\cdot}$$_1$\symbol{40}${\lambda}$$_1$~{\usefont{T1}{pag}{m}{n}{\small{v}}}\symbol{44}~\symbol{40}{\usefont{T1}{pag}{m}{n}{\small{sb}}}~{\usefont{T1}{pag}{m}{n}{\small{v}}}\symbol{41}${\cdot}$$_1$\symbol{91}{\usefont{T1}{pag}{m}{n}{\small{U}}}\symbol{93}${\cdot}$$_1${\usefont{T1}{pag}{m}{n}{\small{le}}}\symbol{39}${\cdot}$$_1${\usefont{T1}{pag}{m}{n}{\small{y}}}\symbol{41}\symbol{93}\symbol{41}\symbol{46}~{\usefont{T1}{pag}{m}{n}{\small{simplify}}}\symbol{46}\\
    \hphantom{ }\hphantom{ }{\usefont{T1}{pag}{m}{n}{\small{{\itshape intros}}}}~{\usefont{T1}{pag}{m}{n}{\small{q}}}\symbol{46}\\
    \hphantom{ }\hphantom{ }{\usefont{T1}{pag}{m}{n}{\small{{\itshape refine}}}}~\symbol{40}{\usefont{T1}{pag}{m}{n}{\small{q}}}${\cdot}$$_0$$\_$\symbol{41}\symbol{46}~{\usefont{T1}{pag}{m}{n}{\small{{\itshape clear}}}}~{\usefont{T1}{pag}{m}{n}{\small{q}}}\symbol{46}\\
    \hphantom{ }\hphantom{ }{\usefont{T1}{pag}{m}{n}{\small{{\itshape unfold}}}}~{\usefont{T1}{pag}{m}{n}{\small{le}}}~{\usefont{T1}{pag}{b}{n}{\small{in}}}~{\usefont{T1}{pag}{m}{n}{\small{h0}}}\symbol{46}~{\usefont{T1}{pag}{m}{n}{\small{simplify\_in}}}~{\usefont{T1}{pag}{m}{n}{\small{h0}}}\symbol{46}\\
    \hphantom{ }\hphantom{ }{\usefont{T1}{pag}{m}{n}{\small{{\itshape unfold}}}}~{\usefont{T1}{pag}{m}{n}{\small{WF}}}~{\usefont{T1}{pag}{b}{n}{\small{in}}}~{\usefont{T1}{pag}{m}{n}{\small{h0}}}\symbol{46}~{\usefont{T1}{pag}{m}{n}{\small{simplify\_in}}}~{\usefont{T1}{pag}{m}{n}{\small{h0}}}\symbol{46}\\
    \hphantom{ }\hphantom{ }{\usefont{T1}{pag}{m}{n}{\small{{\itshape exact}}}}~{\usefont{T1}{pag}{m}{n}{\small{h0}}}\symbol{46}\\
    {\usefont{T1}{pag}{b}{n}{\small{Qed}}}\symbol{46}}}
  \end{displaymath}
  \par
  \begin{displaymath}
    \makebox[1.\linewidth]{\makebox[0.1\linewidth]{}\parbox{0.9\linewidth}{{\usefont{T1}{pag}{b}{n}{\small{Theorem}}}~{\usefont{T1}{pag}{m}{n}{\small{paradox}}}~\symbol{58}~{\usefont{T1}{pag}{m}{n}{\small{El0}}}~{\usefont{T1}{pag}{m}{n}{\small{F}}}\symbol{46}\\
    {\usefont{T1}{pag}{b}{n}{\small{Proof}}}\symbol{46}\\
    \hphantom{ }\hphantom{ }{\usefont{T1}{pag}{m}{n}{\small{{\itshape exact}}}}~\symbol{40}{\usefont{T1}{pag}{m}{n}{\small{lemma2}}}${\cdot}$$_0${\usefont{T1}{pag}{m}{n}{\small{Omega}}}\symbol{41}\symbol{46}\\
    {\usefont{T1}{pag}{b}{n}{\small{Qed}}}\symbol{46}}}
  \end{displaymath}
  \par
  The takeaway insight is that because the paradox does not actually make use of the reduction rules in propositions of {\usefont{T1}{pag}{m}{n}{\small{U0}}}, using equality to model conversion in these propositions doesn't raise any obstacle to the completion of the proof.
  \par
  Nothing in this proof is particularly specific to Coq: it could be done in any variant of Martin-Löf type theory, provided that an identity type is available. Of course, the support of Coq for rewriting significantly helps, if your favourite proof assistant doesn't have a similar feature it may be painful to port this generic paradox.
  \chapter{Applications}
  \par
  In this section we will see a few instances of the generic axiomatisation of Hurkens's proof can help derive contradictions. They come from the file \textsf{theories/Logic/Hurkens.v} of the Coq distribution (version 8.5).
  \par
  \section{Sorts}\label{latex_lib_label_3}
  \par
  A common implementation of universes is to use a sort of the dependent type theory for a universe of $\mbox{\textrm{U}}^-$. In that case. {\usefont{T1}{pag}{m}{n}{\small{El}}} is just the identity.
  \par
  \begin{displaymath}
    \makebox[1.\linewidth]{\makebox[0.1\linewidth]{}\parbox{0.9\linewidth}{{\usefont{T1}{pag}{b}{n}{\small{Variable}}}~{\usefont{T1}{pag}{m}{n}{\small{U}}}~$\coloneqq $~{\usefont{T1}{pag}{b}{n}{\small{Type}}}\symbol{46}\\
    {\usefont{T1}{pag}{b}{n}{\small{Let}}}~{\usefont{T1}{pag}{m}{n}{\small{El}}}~$\coloneqq $~{\usefont{T1}{pag}{b}{n}{\small{fun}}}~{\usefont{T1}{pag}{m}{n}{\small{X}}}~${\Rightarrow}$~{\usefont{T1}{pag}{m}{n}{\small{X}}}\symbol{46}}}
  \end{displaymath}
  \par
  For universes defined this way, small products and their ${\lambda}$-abstraction, application and ${\beta}$-rule are defined straightforwardly ({\usefont{T1}{pag}{m}{n}{\small{eq\_refl}}} is Coq's witness of reflexivity of equality).
  \par
  \begin{displaymath}
    \makebox[1.\linewidth]{\makebox[0.1\linewidth]{}\parbox{0.9\linewidth}{{\usefont{T1}{pag}{b}{n}{\small{Let}}}~{\usefont{T1}{pag}{m}{n}{\small{Forall}}}~\symbol{40}{\usefont{T1}{pag}{m}{n}{\small{A}}}\symbol{58}{\usefont{T1}{pag}{m}{n}{\small{U}}}\symbol{41}~\symbol{40}{\usefont{T1}{pag}{m}{n}{\small{B}}}\symbol{58}{\usefont{T1}{pag}{m}{n}{\small{A}}}~${\rightarrow}$~{\usefont{T1}{pag}{m}{n}{\small{U}}}\symbol{41}~\symbol{58}~{\usefont{T1}{pag}{m}{n}{\small{U}}}~$\coloneqq $~{\usefont{T1}{pag}{b}{n}{\small{forall}}}~{\usefont{T1}{pag}{m}{n}{\small{x}}}\symbol{58}{\usefont{T1}{pag}{m}{n}{\small{A}}}\symbol{44}~{\usefont{T1}{pag}{m}{n}{\small{B}}}~{\usefont{T1}{pag}{m}{n}{\small{x}}}\\
    {\usefont{T1}{pag}{b}{n}{\small{Let}}}~{\usefont{T1}{pag}{m}{n}{\small{lam}}}~{\usefont{T1}{pag}{m}{n}{\small{u}}}~{\usefont{T1}{pag}{m}{n}{\small{B}}}~\symbol{40}{\usefont{T1}{pag}{m}{n}{\small{f}}}\symbol{58}{\usefont{T1}{pag}{b}{n}{\small{forall}}}~{\usefont{T1}{pag}{m}{n}{\small{x}}}\symbol{58}{\usefont{T1}{pag}{m}{n}{\small{A}}}\symbol{44}{\usefont{T1}{pag}{m}{n}{\small{B}}}~{\usefont{T1}{pag}{m}{n}{\small{x}}}\symbol{41}~$\coloneqq $~{\usefont{T1}{pag}{m}{n}{\small{f}}}\\
    {\usefont{T1}{pag}{b}{n}{\small{Let}}}~{\usefont{T1}{pag}{m}{n}{\small{app}}}~{\usefont{T1}{pag}{m}{n}{\small{u}}}~{\usefont{T1}{pag}{m}{n}{\small{B}}}~\symbol{40}{\usefont{T1}{pag}{m}{n}{\small{f}}}\symbol{58}{\usefont{T1}{pag}{b}{n}{\small{forall}}}~{\usefont{T1}{pag}{m}{n}{\small{x}}}\symbol{58}{\usefont{T1}{pag}{m}{n}{\small{A}}}\symbol{44}{\usefont{T1}{pag}{m}{n}{\small{B}}}~{\usefont{T1}{pag}{m}{n}{\small{x}}}\symbol{41}~\symbol{40}{\usefont{T1}{pag}{m}{n}{\small{x}}}\symbol{58}{\usefont{T1}{pag}{m}{n}{\small{A}}}\symbol{41}~$\coloneqq $~{\usefont{T1}{pag}{m}{n}{\small{f}}}~{\usefont{T1}{pag}{m}{n}{\small{x}}}\\
    {\usefont{T1}{pag}{b}{n}{\small{Let}}}~{\usefont{T1}{pag}{m}{n}{\small{beta}}}~{\usefont{T1}{pag}{m}{n}{\small{u}}}~{\usefont{T1}{pag}{m}{n}{\small{B}}}~{\usefont{T1}{pag}{m}{n}{\small{f}}}~{\usefont{T1}{pag}{m}{n}{\small{x}}}~\symbol{58}~{\usefont{T1}{pag}{m}{n}{\small{f}}}~{\usefont{T1}{pag}{m}{n}{\small{x}}}~\symbol{61}~{\usefont{T1}{pag}{m}{n}{\small{f}}}~{\usefont{T1}{pag}{m}{n}{\small{x}}}~$\coloneqq $~{\usefont{T1}{pag}{m}{n}{\small{eq\_refl}}}}}
  \end{displaymath}
  \par
  \section{Impredicative sort}\label{latex_lib_label_4}
  \par
  Impredicativity, for a sort {\usefont{T1}{pag}{m}{n}{\small{U}}}, can also be characterised to some degree. The idea is that there must be a bigger sort {\usefont{T1}{pag}{m}{n}{\small{U}}}\symbol{39} which can be projected onto {\usefont{T1}{pag}{m}{n}{\small{U}}}. See, for example, the bracketing construction in~\cite{HerbelinSpiwack2013}. This projection could be implemented, for instance, for Coq's impredicative {\usefont{T1}{pag}{b}{n}{\small{Prop}}} sort as {\usefont{T1}{pag}{b}{n}{\small{fun}}}~{\usefont{T1}{pag}{m}{n}{\small{X}}}\symbol{58}{\usefont{T1}{pag}{b}{n}{\small{Type}}}~${\Rightarrow}$~{\usefont{T1}{pag}{b}{n}{\small{forall}}}~{\usefont{T1}{pag}{m}{n}{\small{P}}}\symbol{58}{\usefont{T1}{pag}{b}{n}{\small{Prop}}}\symbol{44}~\symbol{40}{\usefont{T1}{pag}{m}{n}{\small{X}}}${\rightarrow}${\usefont{T1}{pag}{m}{n}{\small{P}}}\symbol{41}${\rightarrow}${\usefont{T1}{pag}{m}{n}{\small{P}}}.
  \par
  The signature of Section~\ref{latex_lib_label_3} is extended with the constraint that {\usefont{T1}{pag}{m}{n}{\small{U}}}\symbol{39} is bigger than {\usefont{T1}{pag}{m}{n}{\small{U}}} and a projection.
  \par
  \begin{displaymath}
    \makebox[1.\linewidth]{\makebox[0.1\linewidth]{}\parbox{0.9\linewidth}{{\usefont{T1}{pag}{b}{n}{\small{Let}}}~{\usefont{T1}{pag}{m}{n}{\small{U}}}\symbol{39}~$\coloneqq $~{\usefont{T1}{pag}{b}{n}{\small{Type}}}\symbol{46}\\
    {\usefont{T1}{pag}{b}{n}{\small{Let}}}~{\usefont{T1}{pag}{m}{n}{\small{U}}}\symbol{58}{\usefont{T1}{pag}{m}{n}{\small{U}}}\symbol{39}~$\coloneqq $~{\usefont{T1}{pag}{b}{n}{\small{Type}}}\symbol{46}\\
    {\usefont{T1}{pag}{b}{n}{\small{Variable}}}~{\usefont{T1}{pag}{m}{n}{\small{proj}}}~\symbol{58}~{\usefont{T1}{pag}{m}{n}{\small{U}}}\symbol{39}~${\rightarrow}$~{\usefont{T1}{pag}{m}{n}{\small{U}}}\symbol{46}}}
  \end{displaymath}
  \par
  With the following laws.
  \par
  \begin{displaymath}
    \makebox[1.\linewidth]{\makebox[0.1\linewidth]{}\parbox{0.9\linewidth}{{\usefont{T1}{pag}{b}{n}{\small{Hypothesis}}}~{\usefont{T1}{pag}{m}{n}{\small{proj\_unit}}}~\symbol{58}~{\usefont{T1}{pag}{b}{n}{\small{forall}}}~\symbol{40}{\usefont{T1}{pag}{m}{n}{\small{A}}}\symbol{58}{\usefont{T1}{pag}{m}{n}{\small{U}}}\symbol{39}\symbol{41}\symbol{44}~{\usefont{T1}{pag}{m}{n}{\small{A}}}~${\rightarrow}$~{\usefont{T1}{pag}{m}{n}{\small{proj}}}~{\usefont{T1}{pag}{m}{n}{\small{A}}}\symbol{46}\\
    {\usefont{T1}{pag}{b}{n}{\small{Hypothesis}}}~{\usefont{T1}{pag}{m}{n}{\small{proj\_counit}}}~\symbol{58}~{\usefont{T1}{pag}{b}{n}{\small{forall}}}~\symbol{40}{\usefont{T1}{pag}{m}{n}{\small{F}}}\symbol{58}{\usefont{T1}{pag}{m}{n}{\small{U}}}${\rightarrow}${\usefont{T1}{pag}{m}{n}{\small{U}}}\symbol{41}\symbol{44}~{\usefont{T1}{pag}{m}{n}{\small{proj}}}~\symbol{40}{\usefont{T1}{pag}{b}{n}{\small{forall}}}~{\usefont{T1}{pag}{m}{n}{\small{A}}}\symbol{44}{\usefont{T1}{pag}{m}{n}{\small{F}}}~{\usefont{T1}{pag}{m}{n}{\small{A}}}\symbol{41}~${\rightarrow}$~\symbol{40}{\usefont{T1}{pag}{b}{n}{\small{forall}}}~{\usefont{T1}{pag}{m}{n}{\small{A}}}\symbol{44}{\usefont{T1}{pag}{m}{n}{\small{F}}}~{\usefont{T1}{pag}{m}{n}{\small{A}}}\symbol{41}\symbol{46}\\
    {\usefont{T1}{pag}{b}{n}{\small{Hypothesis}}}~{\usefont{T1}{pag}{m}{n}{\small{proj\_coherent}}}~\symbol{58}~{\usefont{T1}{pag}{b}{n}{\small{forall}}}~\symbol{40}{\usefont{T1}{pag}{m}{n}{\small{F}}}\symbol{58}{\usefont{T1}{pag}{m}{n}{\small{U}}}~${\rightarrow}$~{\usefont{T1}{pag}{m}{n}{\small{U}}}\symbol{41}~\symbol{40}{\usefont{T1}{pag}{m}{n}{\small{f}}}\symbol{58}{\usefont{T1}{pag}{b}{n}{\small{forall}}}~{\usefont{T1}{pag}{m}{n}{\small{x}}}\symbol{58}{\usefont{T1}{pag}{m}{n}{\small{U}}}\symbol{44}~{\usefont{T1}{pag}{m}{n}{\small{F}}}~{\usefont{T1}{pag}{m}{n}{\small{x}}}\symbol{41}~\symbol{40}{\usefont{T1}{pag}{m}{n}{\small{x}}}\symbol{58}{\usefont{T1}{pag}{m}{n}{\small{U}}}\symbol{41}\symbol{44}\\
    \hphantom{ }\hphantom{ }\hphantom{ }\hphantom{ }\hphantom{ }\hphantom{ }\hphantom{ }\hphantom{ }\hphantom{ }\hphantom{ }\hphantom{ }\hphantom{ }\hphantom{ }\hphantom{ }\hphantom{ }\hphantom{ }\hphantom{ }\hphantom{ }\hphantom{ }\hphantom{ }\hphantom{ }\hphantom{ }\hphantom{ }\hphantom{ }\hphantom{ }\hphantom{ }\hphantom{ }\hphantom{ }\hphantom{ }\hphantom{ }{\usefont{T1}{pag}{m}{n}{\small{proj\_counit}}}~$\_$~\symbol{40}{\usefont{T1}{pag}{m}{n}{\small{proj\_unit}}}~$\_$~{\usefont{T1}{pag}{m}{n}{\small{f}}}\symbol{41}~{\usefont{T1}{pag}{m}{n}{\small{x}}}~\symbol{61}~{\usefont{T1}{pag}{m}{n}{\small{f}}}~{\usefont{T1}{pag}{m}{n}{\small{x}}}\symbol{46}}}
  \end{displaymath}
  \par
  The {\usefont{T1}{pag}{m}{n}{\small{proj\_unit}}} law expresses that if {\usefont{T1}{pag}{m}{n}{\small{proj}}} generally diminishes the ability to distinguish between elements of {\usefont{T1}{pag}{m}{n}{\small{A}}}\symbol{58}{\usefont{T1}{pag}{m}{n}{\small{U2}}}, it does not lose elements. We don't have a way back from {\usefont{T1}{pag}{m}{n}{\small{proj}}}~{\usefont{T1}{pag}{m}{n}{\small{A}}} to {\usefont{T1}{pag}{m}{n}{\small{A}}} in general, but {\usefont{T1}{pag}{m}{n}{\small{proj}}} forms a monad. The {\usefont{T1}{pag}{m}{n}{\small{proj\_unit}}} law expresses a small variation on this latter remark.
  \par
  These properties are sufficient to show that {\usefont{T1}{pag}{m}{n}{\small{U}}} is closed by large product. The ${\beta}$-rule, omitted, is easily derived from {\usefont{T1}{pag}{m}{n}{\small{proj\_coherent}}}.
  \par
  \begin{displaymath}
    \makebox[1.\linewidth]{\makebox[0.1\linewidth]{}\parbox{0.9\linewidth}{{\usefont{T1}{pag}{b}{n}{\small{Let}}}~{\usefont{T1}{pag}{m}{n}{\small{ForallU}}}~\symbol{40}{\usefont{T1}{pag}{m}{n}{\small{F}}}\symbol{58}{\usefont{T1}{pag}{m}{n}{\small{U}}}${\rightarrow}${\usefont{T1}{pag}{m}{n}{\small{U}}}\symbol{41}~\symbol{58}~{\usefont{T1}{pag}{m}{n}{\small{U}}}~$\coloneqq $~{\usefont{T1}{pag}{m}{n}{\small{proj}}}~\symbol{40}{\usefont{T1}{pag}{b}{n}{\small{forall}}}~{\usefont{T1}{pag}{m}{n}{\small{A}}}\symbol{44}~{\usefont{T1}{pag}{m}{n}{\small{F}}}~{\usefont{T1}{pag}{m}{n}{\small{A}}}\symbol{41}\symbol{46}\\
    {\usefont{T1}{pag}{b}{n}{\small{Let}}}~{\usefont{T1}{pag}{m}{n}{\small{lamU1}}}~{\usefont{T1}{pag}{m}{n}{\small{F}}}~\symbol{40}{\usefont{T1}{pag}{m}{n}{\small{f}}}\symbol{58}{\usefont{T1}{pag}{b}{n}{\small{forall}}}~{\usefont{T1}{pag}{m}{n}{\small{A}}}\symbol{58}{\usefont{T1}{pag}{m}{n}{\small{U}}}\symbol{44}~{\usefont{T1}{pag}{m}{n}{\small{F}}}~{\usefont{T1}{pag}{m}{n}{\small{A}}}\symbol{41}~\symbol{58}~{\usefont{T1}{pag}{m}{n}{\small{proj}}}\symbol{40}{\usefont{T1}{pag}{b}{n}{\small{forall}}}~{\usefont{T1}{pag}{m}{n}{\small{A}}}\symbol{58}{\usefont{T1}{pag}{m}{n}{\small{U}}}\symbol{44}~{\usefont{T1}{pag}{m}{n}{\small{F}}}~{\usefont{T1}{pag}{m}{n}{\small{A}}}\symbol{41}$\coloneqq $~{\usefont{T1}{pag}{m}{n}{\small{proj\_unit}}}~$\_$~{\usefont{T1}{pag}{m}{n}{\small{f}}}\\
    {\usefont{T1}{pag}{b}{n}{\small{Let}}}~{\usefont{T1}{pag}{m}{n}{\small{appU1}}}~{\usefont{T1}{pag}{m}{n}{\small{F}}}~\symbol{40}{\usefont{T1}{pag}{m}{n}{\small{f}}}\symbol{58}{\usefont{T1}{pag}{m}{n}{\small{proj}}}\symbol{40}{\usefont{T1}{pag}{b}{n}{\small{forall}}}~{\usefont{T1}{pag}{m}{n}{\small{A}}}\symbol{58}{\usefont{T1}{pag}{m}{n}{\small{U}}}\symbol{44}~{\usefont{T1}{pag}{m}{n}{\small{F}}}~{\usefont{T1}{pag}{m}{n}{\small{A}}}\symbol{41}\symbol{41}~\symbol{40}{\usefont{T1}{pag}{m}{n}{\small{A}}}\symbol{58}{\usefont{T1}{pag}{m}{n}{\small{U}}}\symbol{41}~\symbol{58}~{\usefont{T1}{pag}{m}{n}{\small{F}}}~{\usefont{T1}{pag}{m}{n}{\small{A}}}~$\coloneqq $~{\usefont{T1}{pag}{m}{n}{\small{proj\_counit}}}~$\_$~{\usefont{T1}{pag}{m}{n}{\small{f}}}~{\usefont{T1}{pag}{m}{n}{\small{x}}}\symbol{46}}}
  \end{displaymath}
  \par
  We can exploit Coq's universe polymorphism (form version 8.5) to turn this section into a generic definition of impredicative sort. Indeed, under the polymorphic interpretation {\usefont{T1}{pag}{b}{n}{\small{Type}}} represents an arbitrary type, including the impredicative sort {\usefont{T1}{pag}{b}{n}{\small{Prop}}}, which is indeed impredicative in the above sense.
  \par
  \section{Generalising Geuvers's proof}\label{latex_lib_label_5}
  \par
  Geuvers~\cite{Geuvers2007} proves that an impredicative sort {\usefont{T1}{pag}{m}{n}{\small{U1}}} cannot be a retract of an {\usefont{T1}{pag}{m}{n}{\small{U0}}}\symbol{58}{\usefont{T1}{pag}{m}{n}{\small{U1}}}. His proof is made for $\mbox{{\usefont{T1}{pag}{m}{n}{\small{U1}}}}=\mbox{{\usefont{T1}{pag}{b}{n}{\small{Prop}}}}$, but we can instantiate the proof of Section~\ref{latex_lib_label_1} to obtain the same result for any sort which is impredicative sort in the sense of Section~\ref{latex_lib_label_4}.
  \par
  \begin{displaymath}
    \makebox[1.\linewidth]{\makebox[0.1\linewidth]{}\parbox{0.9\linewidth}{{\usefont{T1}{pag}{b}{n}{\small{Let}}}~{\usefont{T1}{pag}{m}{n}{\small{U2}}}~$\coloneqq $~{\usefont{T1}{pag}{b}{n}{\small{Type}}}\symbol{46}\\
    {\usefont{T1}{pag}{b}{n}{\small{Let}}}~{\usefont{T1}{pag}{m}{n}{\small{U1}}}\symbol{58}{\usefont{T1}{pag}{m}{n}{\small{U2}}}~$\coloneqq $~{\usefont{T1}{pag}{b}{n}{\small{Type}}}\symbol{46}\\
    {\usefont{T1}{pag}{b}{n}{\small{Variable}}}~{\usefont{T1}{pag}{m}{n}{\small{U0}}}\symbol{58}{\usefont{T1}{pag}{m}{n}{\small{U1}}}\symbol{46}}}
  \end{displaymath}
  \par
  Where {\usefont{T1}{pag}{m}{n}{\small{U1}}} is impredicative over {\usefont{T1}{pag}{m}{n}{\small{U2}}} as in Section~\ref{latex_lib_label_4}. The retraction is given by the following functions. Only a weak form of retraction is needed were types in {\usefont{T1}{pag}{m}{n}{\small{U1}}} which are ``logically equivalent'' are considered equal.
  \par
  \begin{displaymath}
    \makebox[1.\linewidth]{\makebox[0.1\linewidth]{}\parbox{0.9\linewidth}{{\usefont{T1}{pag}{b}{n}{\small{Variable}}}~{\usefont{T1}{pag}{m}{n}{\small{proj0}}}~\symbol{58}~{\usefont{T1}{pag}{m}{n}{\small{U0}}}~${\rightarrow}$~{\usefont{T1}{pag}{m}{n}{\small{U1}}}\symbol{46}\\
    {\usefont{T1}{pag}{b}{n}{\small{Variable}}}~{\usefont{T1}{pag}{m}{n}{\small{inj0}}}~\symbol{58}~{\usefont{T1}{pag}{m}{n}{\small{U1}}}~${\rightarrow}$~{\usefont{T1}{pag}{m}{n}{\small{U0}}}\symbol{46}\\
    {\usefont{T1}{pag}{b}{n}{\small{Hypothesis}}}~{\usefont{T1}{pag}{m}{n}{\small{inj0\_unit}}}~\symbol{58}~{\usefont{T1}{pag}{b}{n}{\small{forall}}}~\symbol{40}{\usefont{T1}{pag}{m}{n}{\small{b}}}\symbol{58}{\usefont{T1}{pag}{m}{n}{\small{U1}}}\symbol{41}\symbol{44}~{\usefont{T1}{pag}{m}{n}{\small{b}}}~${\rightarrow}$~{\usefont{T1}{pag}{m}{n}{\small{proj0}}}~\symbol{40}{\usefont{T1}{pag}{m}{n}{\small{inj0}}}~{\usefont{T1}{pag}{m}{n}{\small{b}}}\symbol{41}\symbol{46}\\
    {\usefont{T1}{pag}{b}{n}{\small{Hypothesis}}}~{\usefont{T1}{pag}{m}{n}{\small{inj0\_counit}}}~\symbol{58}~{\usefont{T1}{pag}{b}{n}{\small{forall}}}~\symbol{40}{\usefont{T1}{pag}{m}{n}{\small{b}}}\symbol{58}{\usefont{T1}{pag}{m}{n}{\small{U1}}}\symbol{41}\symbol{44}~{\usefont{T1}{pag}{m}{n}{\small{proj0}}}~\symbol{40}{\usefont{T1}{pag}{m}{n}{\small{inj0}}}~{\usefont{T1}{pag}{m}{n}{\small{b}}}\symbol{41}~${\rightarrow}$~{\usefont{T1}{pag}{m}{n}{\small{b}}}\symbol{46}}}
  \end{displaymath}
  \par
  From this (weak) retraction we can define {\usefont{T1}{pag}{m}{n}{\small{El0}}} and corresponding products for {\usefont{T1}{pag}{m}{n}{\small{U0}}} despite the fact that {\usefont{T1}{pag}{m}{n}{\small{U0}}} is not necessarily a sort.
  \par
  \begin{displaymath}
    \makebox[1.\linewidth]{\makebox[0.1\linewidth]{}\parbox{0.9\linewidth}{{\usefont{T1}{pag}{b}{n}{\small{Let}}}~{\usefont{T1}{pag}{m}{n}{\small{El0}}}~\symbol{40}{\usefont{T1}{pag}{m}{n}{\small{u}}}\symbol{58}{\usefont{T1}{pag}{m}{n}{\small{U0}}}\symbol{41}~$\coloneqq $~{\usefont{T1}{pag}{m}{n}{\small{proj0}}}~{\usefont{T1}{pag}{m}{n}{\small{u}}}\\
    {\usefont{T1}{pag}{b}{n}{\small{Let}}}~{\usefont{T1}{pag}{m}{n}{\small{Forall0}}}~\symbol{40}{\usefont{T1}{pag}{m}{n}{\small{u}}}\symbol{58}{\usefont{T1}{pag}{m}{n}{\small{U0}}}\symbol{41}~\symbol{40}{\usefont{T1}{pag}{m}{n}{\small{B}}}\symbol{58}{\usefont{T1}{pag}{m}{n}{\small{proj0}}}~{\usefont{T1}{pag}{m}{n}{\small{u}}}~${\rightarrow}$~{\usefont{T1}{pag}{m}{n}{\small{U0}}}\symbol{41}~\symbol{58}~{\usefont{T1}{pag}{m}{n}{\small{U0}}}~$\coloneqq $~{\usefont{T1}{pag}{m}{n}{\small{inj0}}}~\symbol{40}{\usefont{T1}{pag}{b}{n}{\small{forall}}}~{\usefont{T1}{pag}{m}{n}{\small{x}}}\symbol{58}{\usefont{T1}{pag}{m}{n}{\small{proj0}}}~{\usefont{T1}{pag}{m}{n}{\small{u}}}\symbol{44}~{\usefont{T1}{pag}{m}{n}{\small{proj0}}}~\symbol{40}{\usefont{T1}{pag}{m}{n}{\small{B}}}~{\usefont{T1}{pag}{m}{n}{\small{x}}}\symbol{41}\symbol{41}\\
    {\usefont{T1}{pag}{b}{n}{\small{Let}}}~{\usefont{T1}{pag}{m}{n}{\small{Lambda0}}}~{\usefont{T1}{pag}{m}{n}{\small{u}}}~{\usefont{T1}{pag}{m}{n}{\small{B}}}~\symbol{40}{\usefont{T1}{pag}{m}{n}{\small{f}}}\symbol{58}{\usefont{T1}{pag}{b}{n}{\small{forall}}}~{\usefont{T1}{pag}{m}{n}{\small{x}}}\symbol{58}{\usefont{T1}{pag}{m}{n}{\small{proj0}}}~{\usefont{T1}{pag}{m}{n}{\small{u}}}\symbol{44}~{\usefont{T1}{pag}{m}{n}{\small{proj0}}}~\symbol{40}{\usefont{T1}{pag}{m}{n}{\small{B}}}~{\usefont{T1}{pag}{m}{n}{\small{x}}}\symbol{41}\symbol{41}\\
    \hphantom{ }\hphantom{ }\hphantom{ }\symbol{58}~{\usefont{T1}{pag}{m}{n}{\small{proj0}}}~\symbol{40}{\usefont{T1}{pag}{m}{n}{\small{inj0}}}~\symbol{40}{\usefont{T1}{pag}{b}{n}{\small{forall}}}~{\usefont{T1}{pag}{m}{n}{\small{x}}}\symbol{58}{\usefont{T1}{pag}{m}{n}{\small{proj0}}}~{\usefont{T1}{pag}{m}{n}{\small{u}}}\symbol{44}~{\usefont{T1}{pag}{m}{n}{\small{proj0}}}~\symbol{40}{\usefont{T1}{pag}{m}{n}{\small{B}}}~{\usefont{T1}{pag}{m}{n}{\small{x}}}\symbol{41}\symbol{41}\symbol{41}~$\coloneqq $~{\usefont{T1}{pag}{m}{n}{\small{inj0\_unit}}}~$\_$~{\usefont{T1}{pag}{m}{n}{\small{f}}}\symbol{46}\\
    {\usefont{T1}{pag}{b}{n}{\small{Let}}}~{\usefont{T1}{pag}{m}{n}{\small{app0}}}~{\usefont{T1}{pag}{b}{n}{\small{forall}}}~{\usefont{T1}{pag}{m}{n}{\small{u}}}~{\usefont{T1}{pag}{m}{n}{\small{B}}}~\symbol{40}{\usefont{T1}{pag}{m}{n}{\small{f}}}\symbol{58}{\usefont{T1}{pag}{m}{n}{\small{proj0}}}~\symbol{40}{\usefont{T1}{pag}{m}{n}{\small{inj0}}}~\symbol{40}{\usefont{T1}{pag}{b}{n}{\small{forall}}}~{\usefont{T1}{pag}{m}{n}{\small{x}}}\symbol{58}{\usefont{T1}{pag}{m}{n}{\small{proj0}}}~{\usefont{T1}{pag}{m}{n}{\small{u}}}\symbol{44}~{\usefont{T1}{pag}{m}{n}{\small{proj0}}}~\symbol{40}{\usefont{T1}{pag}{m}{n}{\small{B}}}~{\usefont{T1}{pag}{m}{n}{\small{x}}}\symbol{41}\symbol{41}\symbol{41}\symbol{41}~\symbol{40}{\usefont{T1}{pag}{m}{n}{\small{x}}}\symbol{58}{\usefont{T1}{pag}{m}{n}{\small{proj0}}}~{\usefont{T1}{pag}{m}{n}{\small{u}}}\symbol{41}\\
    \hphantom{ }\hphantom{ }\hphantom{ }\symbol{58}~{\usefont{T1}{pag}{m}{n}{\small{proj0}}}~\symbol{40}{\usefont{T1}{pag}{m}{n}{\small{B}}}~{\usefont{T1}{pag}{m}{n}{\small{x}}}\symbol{41}~$\coloneqq $~{\usefont{T1}{pag}{m}{n}{\small{inj0\_counit}}}~$\_$~{\usefont{T1}{pag}{m}{n}{\small{f}}}~{\usefont{T1}{pag}{m}{n}{\small{x}}}}}
  \end{displaymath}
  \par
  Large products are define much the same:
  \begin{displaymath}
    \makebox[1.\linewidth]{\makebox[0.1\linewidth]{}\parbox{0.9\linewidth}{{\usefont{T1}{pag}{b}{n}{\small{Let}}}~{\usefont{T1}{pag}{m}{n}{\small{Forall0}}}~\symbol{40}{\usefont{T1}{pag}{m}{n}{\small{u}}}\symbol{58}{\usefont{T1}{pag}{m}{n}{\small{U1}}}\symbol{41}~\symbol{40}{\usefont{T1}{pag}{m}{n}{\small{B}}}\symbol{58}{\usefont{T1}{pag}{m}{n}{\small{u}}}~${\rightarrow}$~{\usefont{T1}{pag}{m}{n}{\small{U0}}}\symbol{41}~\symbol{58}~{\usefont{T1}{pag}{m}{n}{\small{U0}}}~$\coloneqq $~{\usefont{T1}{pag}{m}{n}{\small{inj0}}}~\symbol{40}{\usefont{T1}{pag}{b}{n}{\small{forall}}}~{\usefont{T1}{pag}{m}{n}{\small{x}}}\symbol{58}{\usefont{T1}{pag}{m}{n}{\small{u}}}\symbol{44}~{\usefont{T1}{pag}{m}{n}{\small{proj0}}}~\symbol{40}{\usefont{T1}{pag}{m}{n}{\small{B}}}~{\usefont{T1}{pag}{m}{n}{\small{x}}}\symbol{41}\symbol{41}\\
    {\usefont{T1}{pag}{b}{n}{\small{Let}}}~{\usefont{T1}{pag}{m}{n}{\small{Lambda0}}}~{\usefont{T1}{pag}{m}{n}{\small{u}}}~{\usefont{T1}{pag}{m}{n}{\small{B}}}~\symbol{40}{\usefont{T1}{pag}{m}{n}{\small{f}}}\symbol{58}{\usefont{T1}{pag}{b}{n}{\small{forall}}}~{\usefont{T1}{pag}{m}{n}{\small{x}}}\symbol{58}{\usefont{T1}{pag}{m}{n}{\small{u}}}\symbol{44}~{\usefont{T1}{pag}{m}{n}{\small{proj0}}}~\symbol{40}{\usefont{T1}{pag}{m}{n}{\small{B}}}~{\usefont{T1}{pag}{m}{n}{\small{x}}}\symbol{41}\symbol{41}\\
    \hphantom{ }\hphantom{ }\hphantom{ }\symbol{58}~{\usefont{T1}{pag}{m}{n}{\small{proj0}}}~\symbol{40}{\usefont{T1}{pag}{m}{n}{\small{inj0}}}~\symbol{40}{\usefont{T1}{pag}{b}{n}{\small{forall}}}~{\usefont{T1}{pag}{m}{n}{\small{x}}}\symbol{58}{\usefont{T1}{pag}{m}{n}{\small{u}}}\symbol{44}~{\usefont{T1}{pag}{m}{n}{\small{proj0}}}~\symbol{40}{\usefont{T1}{pag}{m}{n}{\small{B}}}~{\usefont{T1}{pag}{m}{n}{\small{x}}}\symbol{41}\symbol{41}\symbol{41}~$\coloneqq $~{\usefont{T1}{pag}{m}{n}{\small{inj0\_unit}}}~$\_$~{\usefont{T1}{pag}{m}{n}{\small{f}}}\symbol{46}\\
    {\usefont{T1}{pag}{b}{n}{\small{Let}}}~{\usefont{T1}{pag}{m}{n}{\small{app0}}}~{\usefont{T1}{pag}{b}{n}{\small{forall}}}~{\usefont{T1}{pag}{m}{n}{\small{u}}}~{\usefont{T1}{pag}{m}{n}{\small{B}}}~\symbol{40}{\usefont{T1}{pag}{m}{n}{\small{f}}}\symbol{58}{\usefont{T1}{pag}{m}{n}{\small{proj0}}}~\symbol{40}{\usefont{T1}{pag}{m}{n}{\small{inj0}}}~\symbol{40}{\usefont{T1}{pag}{b}{n}{\small{forall}}}~{\usefont{T1}{pag}{m}{n}{\small{x}}}\symbol{58}{\usefont{T1}{pag}{m}{n}{\small{u}}}\symbol{44}~{\usefont{T1}{pag}{m}{n}{\small{proj0}}}~\symbol{40}{\usefont{T1}{pag}{m}{n}{\small{B}}}~{\usefont{T1}{pag}{m}{n}{\small{x}}}\symbol{41}\symbol{41}\symbol{41}\symbol{41}~\symbol{40}{\usefont{T1}{pag}{m}{n}{\small{x}}}\symbol{58}{\usefont{T1}{pag}{m}{n}{\small{u}}}\symbol{41}\\
    \hphantom{ }\hphantom{ }\hphantom{ }\symbol{58}~{\usefont{T1}{pag}{m}{n}{\small{proj0}}}~\symbol{40}{\usefont{T1}{pag}{m}{n}{\small{B}}}~{\usefont{T1}{pag}{m}{n}{\small{x}}}\symbol{41}~$\coloneqq $~{\usefont{T1}{pag}{m}{n}{\small{inj0\_counit}}}~$\_$~{\usefont{T1}{pag}{m}{n}{\small{f}}}~{\usefont{T1}{pag}{m}{n}{\small{x}}}}}
  \end{displaymath}
  \par
  From this, the paradox is set up, so we can deduce that every proposition of {\usefont{T1}{pag}{m}{n}{\small{P}}}\symbol{58}{\usefont{T1}{pag}{m}{n}{\small{U0}}} is ``inhabited'' in that $\mbox{{\usefont{T1}{pag}{m}{n}{\small{El0}}}~{\usefont{T1}{pag}{m}{n}{\small{P}}}}=\mbox{{\usefont{T1}{pag}{m}{n}{\small{proj0}}}~{\usefont{T1}{pag}{m}{n}{\small{P}}}}$ is inhabited, and therefore, that every proposition of {\usefont{T1}{pag}{m}{n}{\small{F}}}\symbol{58}{\usefont{T1}{pag}{m}{n}{\small{U1}}} is inhabited since {\usefont{T1}{pag}{m}{n}{\small{inj0}}}~{\usefont{T1}{pag}{m}{n}{\small{F}}}\symbol{58}{\usefont{T1}{pag}{m}{n}{\small{U0}}} is ``inhabited'' in the sense of {\usefont{T1}{pag}{m}{n}{\small{U0}}}, \emph{i.e.} {\usefont{T1}{pag}{m}{n}{\small{proj0}}}~\symbol{40}{\usefont{T1}{pag}{m}{n}{\small{inj0}}}~{\usefont{T1}{pag}{m}{n}{\small{F}}}\symbol{41} is inhabited, then {\usefont{T1}{pag}{m}{n}{\small{inj0\_counit}}} concludes.
  \par
  Since {\usefont{T1}{pag}{b}{n}{\small{Prop}}} is an instance of the signature of Section~\ref{latex_lib_label_4}, we prove, like Geuvers, that {\usefont{T1}{pag}{b}{n}{\small{Prop}}} is not a retract of a proposition {\usefont{T1}{pag}{m}{n}{\small{P}}}\symbol{58}{\usefont{T1}{pag}{b}{n}{\small{Prop}}}.
  \par
  \section{Excluded middle and proof irrelevance}\label{latex_lib_label_6}
  \par
    Geuvers proof, from Section~\ref{latex_lib_label_5}, helps proving a result, by Coquand~\cite{Coquand1989}, that excluded middle, in an impredicative sort makes it proof irrelevant, \emph{i.e.} every type in that sort have at most one element. This proof appear in the Coq distribution in the file \textsf{theories/Logic/ClassicalFact.v}, presumably written by Hugo Herbelin. It uses Geuvers result and was mostly unmodified with the new proof of said result. With the characterisation of Section~\ref{latex_lib_label_4}, this could be done in an arbitrary impredicative sort, but the Coq proof is done only for the impredicative sort {\usefont{T1}{pag}{b}{n}{\small{Prop}}}, and we will present it that way for simplicity.
  \par
    The basic idea is that excluded middle:
    \begin{displaymath}
    \makebox[1.\linewidth]{\makebox[0.1\linewidth]{}\parbox{0.9\linewidth}{{\usefont{T1}{pag}{b}{n}{\small{Variable}}}~{\usefont{T1}{pag}{m}{n}{\small{em}}}\symbol{58}~{\usefont{T1}{pag}{b}{n}{\small{forall}}}~{\usefont{T1}{pag}{m}{n}{\small{A}}}\symbol{58}{\usefont{T1}{pag}{b}{n}{\small{Prop}}}\symbol{44}~{\usefont{T1}{pag}{m}{n}{\small{A}}}${\lor}$${\lnot}${\usefont{T1}{pag}{m}{n}{\small{A}}}\symbol{46}}}
  \end{displaymath}
  turns the {\usefont{T1}{pag}{b}{n}{\small{Prop}}} sort into a boolean universe with only two elements. So assuming a proposition with two \emph{distinct} values
    \begin{displaymath}
    \makebox[1.\linewidth]{\makebox[0.1\linewidth]{}\parbox{0.9\linewidth}{{\usefont{T1}{pag}{b}{n}{\small{Variable}}}~{\usefont{T1}{pag}{m}{n}{\small{U0}}}\symbol{58}{\usefont{T1}{pag}{b}{n}{\small{Prop}}}\symbol{46}\\
    {\usefont{T1}{pag}{b}{n}{\small{Variables}}}~{\usefont{T1}{pag}{m}{n}{\small{t}}}~{\usefont{T1}{pag}{m}{n}{\small{f}}}~\symbol{58}~{\usefont{T1}{pag}{m}{n}{\small{U0}}}\symbol{46}\\
    {\usefont{T1}{pag}{b}{n}{\small{Hypothesis}}}~{\usefont{T1}{pag}{m}{n}{\small{not\_eq\_t\_f}}}~\symbol{58}~{\usefont{T1}{pag}{m}{n}{\small{t}}}~${\ne}$~{\usefont{T1}{pag}{m}{n}{\small{f}}}\symbol{46}}}
  \end{displaymath}
  we can reflect {\usefont{T1}{pag}{b}{n}{\small{Prop}}} into {\usefont{T1}{pag}{m}{n}{\small{U0}}} proposition as in Section~\ref{latex_lib_label_5}. Where {\usefont{T1}{pag}{m}{n}{\small{True}}} is reflected as {\usefont{T1}{pag}{m}{n}{\small{t}}} and {\usefont{T1}{pag}{m}{n}{\small{False}}} as {\usefont{T1}{pag}{m}{n}{\small{f}}}, as the names suggest.
  \par
    This is formalised as a retraction given by:
    \begin{displaymath}
    \makebox[1.\linewidth]{\makebox[0.1\linewidth]{}\parbox{0.9\linewidth}{{\usefont{T1}{pag}{b}{n}{\small{Let}}}~{\usefont{T1}{pag}{m}{n}{\small{inj0}}}~\symbol{40}{\usefont{T1}{pag}{m}{n}{\small{A}}}\symbol{58}{\usefont{T1}{pag}{b}{n}{\small{Prop}}}\symbol{41}~\symbol{58}~{\usefont{T1}{pag}{m}{n}{\small{U0}}}~$\coloneqq $~{\usefont{T1}{pag}{m}{n}{\small{or\_ind}}}~{\usefont{T1}{pag}{m}{n}{\small{A}}}~\symbol{40}${\lnot}${\usefont{T1}{pag}{m}{n}{\small{A}}}\symbol{41}~{\usefont{T1}{pag}{m}{n}{\small{U0}}}~\symbol{40}{\usefont{T1}{pag}{b}{n}{\small{fun}}}~$\_$~${\Rightarrow}$~{\usefont{T1}{pag}{m}{n}{\small{t}}}\symbol{41}~\symbol{40}{\usefont{T1}{pag}{b}{n}{\small{fun}}}~$\_$~${\Rightarrow}$~{\usefont{T1}{pag}{m}{n}{\small{f}}}\symbol{41}~\symbol{40}{\usefont{T1}{pag}{m}{n}{\small{em}}}~{\usefont{T1}{pag}{m}{n}{\small{A}}}\symbol{41}\symbol{46}\\
    {\usefont{T1}{pag}{b}{n}{\small{Let}}}~{\usefont{T1}{pag}{m}{n}{\small{proj0}}}~\symbol{40}{\usefont{T1}{pag}{m}{n}{\small{x}}}\symbol{58}{\usefont{T1}{pag}{m}{n}{\small{U0}}}\symbol{41}~\symbol{58}~{\usefont{T1}{pag}{b}{n}{\small{Prop}}}~$\coloneqq $~{\usefont{T1}{pag}{m}{n}{\small{t}}}~\symbol{61}~{\usefont{T1}{pag}{m}{n}{\small{x}}}\symbol{46}}}
  \end{displaymath}
    Where {\usefont{T1}{pag}{m}{n}{\small{or\_ind}}}\symbol{58}{\usefont{T1}{pag}{b}{n}{\small{forall}}}~{\usefont{T1}{pag}{m}{n}{\small{A}}}~{\usefont{T1}{pag}{m}{n}{\small{B}}}~{\usefont{T1}{pag}{m}{n}{\small{P}}}~\symbol{58}~{\usefont{T1}{pag}{b}{n}{\small{Prop}}}\symbol{44}~\symbol{40}{\usefont{T1}{pag}{m}{n}{\small{A}}}~${\rightarrow}$~{\usefont{T1}{pag}{m}{n}{\small{P}}}\symbol{41}~${\rightarrow}$~\symbol{40}{\usefont{T1}{pag}{m}{n}{\small{B}}}~${\rightarrow}$~{\usefont{T1}{pag}{m}{n}{\small{P}}}\symbol{41}~${\rightarrow}$~{\usefont{T1}{pag}{m}{n}{\small{A}}}~${\lor}$~{\usefont{T1}{pag}{m}{n}{\small{B}}}~${\rightarrow}$~{\usefont{T1}{pag}{m}{n}{\small{P}}} is the elimination principle of disjunction.
  \par
  We are left to prove the unit and counit laws of {\usefont{T1}{pag}{m}{n}{\small{inj0}}} and {\usefont{T1}{pag}{m}{n}{\small{proj0}}} to satisfy the premisses of the paradox in Section~\ref{latex_lib_label_5}. The unit law is direct:
  \begin{displaymath}
    \makebox[1.\linewidth]{\makebox[0.1\linewidth]{}\parbox{0.9\linewidth}{{\usefont{T1}{pag}{b}{n}{\small{Lemma}}}~{\usefont{T1}{pag}{m}{n}{\small{inj0\_unit}}}~\symbol{58}~{\usefont{T1}{pag}{b}{n}{\small{forall}}}~{\usefont{T1}{pag}{m}{n}{\small{A}}}\symbol{58}{\usefont{T1}{pag}{b}{n}{\small{Prop}}}\symbol{44}~{\usefont{T1}{pag}{m}{n}{\small{A}}}~${\rightarrow}$~{\usefont{T1}{pag}{m}{n}{\small{proj0}}}~\symbol{40}{\usefont{T1}{pag}{m}{n}{\small{inj0}}}~{\usefont{T1}{pag}{m}{n}{\small{A}}}\symbol{41}\symbol{46}\\
    {\usefont{T1}{pag}{b}{n}{\small{Proof}}}\symbol{46}\\
    \hphantom{ }\hphantom{ }{\usefont{T1}{pag}{m}{n}{\small{{\itshape intros}}}}~{\usefont{T1}{pag}{m}{n}{\small{A}}}~{\usefont{T1}{pag}{m}{n}{\small{x}}}\symbol{46}~{\usefont{T1}{pag}{m}{n}{\small{{\itshape unfold}}}}~{\usefont{T1}{pag}{m}{n}{\small{proj0}}}\symbol{44}~{\usefont{T1}{pag}{m}{n}{\small{inj0}}}\symbol{46}\\
    \hphantom{ }\hphantom{ }{\usefont{T1}{pag}{m}{n}{\small{{\itshape destruct}}}}~\symbol{40}{\usefont{T1}{pag}{m}{n}{\small{em}}}~{\usefont{T1}{pag}{m}{n}{\small{A}}}\symbol{41}~{\usefont{T1}{pag}{m}{n}{\small{{\itshape as}}}}~\symbol{91}{\usefont{T1}{pag}{m}{n}{\small{h}}}~\hspace{0.1em}\rule[-0.6ex]{1.sp}{1.\baselineskip}~{\usefont{T1}{pag}{m}{n}{\small{h}}}\symbol{93}\symbol{46}\\
    \hphantom{ }\hphantom{ }\symbol{43}~{\usefont{T1}{pag}{m}{n}{\small{{\itshape reflexivity}}}}\symbol{46}\\
    \hphantom{ }\hphantom{ }\symbol{43}~{\usefont{T1}{pag}{m}{n}{\small{{\itshape contradiction}}}}\symbol{46}\\
    {\usefont{T1}{pag}{b}{n}{\small{Qed}}}\symbol{46}}}
  \end{displaymath}
    The counit law is the step that makes a crucial use of the {\usefont{T1}{pag}{m}{n}{\small{not\_eq\_t\_f}}} hypothesis:
  \begin{displaymath}
    \makebox[1.\linewidth]{\makebox[0.1\linewidth]{}\parbox{0.9\linewidth}{{\usefont{T1}{pag}{b}{n}{\small{Lemma}}}~{\usefont{T1}{pag}{m}{n}{\small{inj0\_counit}}}~\symbol{58}~{\usefont{T1}{pag}{b}{n}{\small{forall}}}~{\usefont{T1}{pag}{m}{n}{\small{A}}}\symbol{58}{\usefont{T1}{pag}{b}{n}{\small{Prop}}}\symbol{44}~{\usefont{T1}{pag}{m}{n}{\small{proj0}}}~\symbol{40}{\usefont{T1}{pag}{m}{n}{\small{inj0}}}~{\usefont{T1}{pag}{m}{n}{\small{A}}}\symbol{41}~${\rightarrow}$~{\usefont{T1}{pag}{m}{n}{\small{A}}}\symbol{46}\\
    {\usefont{T1}{pag}{b}{n}{\small{Proof}}}\symbol{46}\\
    \hphantom{ }\hphantom{ }{\usefont{T1}{pag}{m}{n}{\small{{\itshape intros}}}}~{\usefont{T1}{pag}{m}{n}{\small{A}}}~{\usefont{T1}{pag}{m}{n}{\small{h}}}\symbol{46}~{\usefont{T1}{pag}{m}{n}{\small{{\itshape unfold}}}}~{\usefont{T1}{pag}{m}{n}{\small{proj0}}}\symbol{44}~{\usefont{T1}{pag}{m}{n}{\small{inj0}}}~{\usefont{T1}{pag}{b}{n}{\small{in}}}~$*$\symbol{46}\\
    \hphantom{ }\hphantom{ }{\usefont{T1}{pag}{m}{n}{\small{{\itshape destruct}}}}~\symbol{40}{\usefont{T1}{pag}{m}{n}{\small{em}}}~{\usefont{T1}{pag}{m}{n}{\small{A}}}\symbol{41}~{\usefont{T1}{pag}{m}{n}{\small{{\itshape as}}}}~\symbol{91}{\usefont{T1}{pag}{m}{n}{\small{l}}}~\hspace{0.1em}\rule[-0.6ex]{1.sp}{1.\baselineskip}~{\usefont{T1}{pag}{m}{n}{\small{l}}}\symbol{93}\symbol{46}\\
    \hphantom{ }\hphantom{ }\symbol{43}~{\usefont{T1}{pag}{m}{n}{\small{{\itshape apply}}}}~{\usefont{T1}{pag}{m}{n}{\small{l}}}\symbol{46}\\
    \hphantom{ }\hphantom{ }\symbol{43}~{\usefont{T1}{pag}{m}{n}{\small{absurd}}}~\symbol{40}{\usefont{T1}{pag}{m}{n}{\small{t}}}\symbol{61}{\usefont{T1}{pag}{m}{n}{\small{f}}}\symbol{41}\symbol{46}\\
    \hphantom{ }\hphantom{ }\hphantom{ }\hphantom{ }$*$~{\usefont{T1}{pag}{m}{n}{\small{{\itshape apply}}}}~{\usefont{T1}{pag}{m}{n}{\small{not\_eq\_t\_f}}}\symbol{46}\\
    \hphantom{ }\hphantom{ }\hphantom{ }\hphantom{ }$*$~{\usefont{T1}{pag}{m}{n}{\small{{\itshape apply}}}}~{\usefont{T1}{pag}{m}{n}{\small{h}}}\symbol{46}\\
    {\usefont{T1}{pag}{b}{n}{\small{Qed}}}\symbol{46}}}
  \end{displaymath}
  \par
    Section~\ref{latex_lib_label_5} then yields a contradiction. Since {\usefont{T1}{pag}{m}{n}{\small{U0}}} is arbitrary we have: {\usefont{T1}{pag}{b}{n}{\small{forall}}}~\symbol{40}{\usefont{T1}{pag}{m}{n}{\small{A}}}\symbol{58}{\usefont{T1}{pag}{b}{n}{\small{Prop}}}\symbol{41}~\symbol{40}{\usefont{T1}{pag}{m}{n}{\small{x}}}~{\usefont{T1}{pag}{m}{n}{\small{y}}}\symbol{58}{\usefont{T1}{pag}{m}{n}{\small{A}}}\symbol{41}\symbol{44}~${\lnot}$${\lnot}$\symbol{40}{\usefont{T1}{pag}{m}{n}{\small{x}}}\symbol{61}{\usefont{T1}{pag}{m}{n}{\small{y}}}\symbol{41}. A last application of the excluded middle yields the expected result:
    \begin{displaymath}
    \makebox[1.\linewidth]{\makebox[0.1\linewidth]{}\parbox{0.9\linewidth}{{\usefont{T1}{pag}{b}{n}{\small{forall}}}~\symbol{40}{\usefont{T1}{pag}{m}{n}{\small{A}}}\symbol{58}{\usefont{T1}{pag}{b}{n}{\small{Prop}}}\symbol{41}~\symbol{40}{\usefont{T1}{pag}{m}{n}{\small{x}}}~{\usefont{T1}{pag}{m}{n}{\small{y}}}\symbol{58}{\usefont{T1}{pag}{m}{n}{\small{A}}}\symbol{41}\symbol{44}~{\usefont{T1}{pag}{m}{n}{\small{x}}}\symbol{61}{\usefont{T1}{pag}{m}{n}{\small{y}}}}}
  \end{displaymath}
  \par
  \section{Variants of Prop}\label{latex_lib_label_7}
  \par
    A (monadic) modality on {\usefont{T1}{pag}{b}{n}{\small{Prop}}} is given by a mapping:
    \begin{displaymath}
    \makebox[1.\linewidth]{\makebox[0.1\linewidth]{}\parbox{0.9\linewidth}{{\usefont{T1}{pag}{b}{n}{\small{Variable}}}~{\usefont{T1}{pag}{m}{n}{\small{M}}}~\symbol{58}~{\usefont{T1}{pag}{b}{n}{\small{Prop}}}~${\rightarrow}$~{\usefont{T1}{pag}{b}{n}{\small{Prop}}}\symbol{46}}}
  \end{displaymath}
    Together with the following laws:
  \begin{displaymath}
    \makebox[1.\linewidth]{\makebox[0.1\linewidth]{}\parbox{0.9\linewidth}{{\usefont{T1}{pag}{b}{n}{\small{Hypothesis}}}~{\usefont{T1}{pag}{m}{n}{\small{unit}}}~\symbol{58}~{\usefont{T1}{pag}{b}{n}{\small{forall}}}~{\usefont{T1}{pag}{m}{n}{\small{A}}}\symbol{58}{\usefont{T1}{pag}{b}{n}{\small{Prop}}}\symbol{44}~{\usefont{T1}{pag}{m}{n}{\small{A}}}~${\rightarrow}$~{\usefont{T1}{pag}{m}{n}{\small{M}}}~{\usefont{T1}{pag}{m}{n}{\small{A}}}\symbol{46}\\
    {\usefont{T1}{pag}{b}{n}{\small{Hypothesis}}}~{\usefont{T1}{pag}{m}{n}{\small{join}}}~\symbol{58}~{\usefont{T1}{pag}{b}{n}{\small{forall}}}~{\usefont{T1}{pag}{m}{n}{\small{A}}}\symbol{58}{\usefont{T1}{pag}{b}{n}{\small{Prop}}}\symbol{44}~{\usefont{T1}{pag}{m}{n}{\small{M}}}~\symbol{40}{\usefont{T1}{pag}{m}{n}{\small{M}}}~{\usefont{T1}{pag}{m}{n}{\small{A}}}\symbol{41}~${\rightarrow}$~{\usefont{T1}{pag}{m}{n}{\small{M}}}~{\usefont{T1}{pag}{m}{n}{\small{A}}}\symbol{46}\\
    {\usefont{T1}{pag}{b}{n}{\small{Hypothesis}}}~{\usefont{T1}{pag}{m}{n}{\small{incr}}}~\symbol{58}~{\usefont{T1}{pag}{b}{n}{\small{forall}}}~{\usefont{T1}{pag}{m}{n}{\small{A}}}~{\usefont{T1}{pag}{m}{n}{\small{B}}}\symbol{58}{\usefont{T1}{pag}{b}{n}{\small{Prop}}}\symbol{44}~\symbol{40}{\usefont{T1}{pag}{m}{n}{\small{A}}}${\rightarrow}${\usefont{T1}{pag}{m}{n}{\small{B}}}\symbol{41}~${\rightarrow}$~{\usefont{T1}{pag}{m}{n}{\small{M}}}~{\usefont{T1}{pag}{m}{n}{\small{A}}}~${\rightarrow}$~{\usefont{T1}{pag}{m}{n}{\small{M}}}~{\usefont{T1}{pag}{m}{n}{\small{B}}}\symbol{46}}}
  \end{displaymath}
    Such a modality is automatically equipped with a distribution property over arbitrary conjunctions:
  \begin{displaymath}
    \makebox[1.\linewidth]{\makebox[0.1\linewidth]{}\parbox{0.9\linewidth}{{\usefont{T1}{pag}{b}{n}{\small{Lemma}}}~{\usefont{T1}{pag}{m}{n}{\small{strength}}}\symbol{58}~{\usefont{T1}{pag}{b}{n}{\small{forall}}}~{\usefont{T1}{pag}{m}{n}{\small{A}}}~\symbol{40}{\usefont{T1}{pag}{m}{n}{\small{P}}}\symbol{58}{\usefont{T1}{pag}{m}{n}{\small{A}}}${\rightarrow}${\usefont{T1}{pag}{b}{n}{\small{Prop}}}\symbol{41}\symbol{44}~{\usefont{T1}{pag}{m}{n}{\small{M}}}\symbol{40}{\usefont{T1}{pag}{b}{n}{\small{forall}}}~{\usefont{T1}{pag}{m}{n}{\small{x}}}\symbol{58}{\usefont{T1}{pag}{m}{n}{\small{A}}}\symbol{44}{\usefont{T1}{pag}{m}{n}{\small{P}}}~{\usefont{T1}{pag}{m}{n}{\small{x}}}\symbol{41}~${\rightarrow}$~{\usefont{T1}{pag}{b}{n}{\small{forall}}}~{\usefont{T1}{pag}{m}{n}{\small{x}}}\symbol{58}{\usefont{T1}{pag}{m}{n}{\small{A}}}\symbol{44}{\usefont{T1}{pag}{m}{n}{\small{M}}}\symbol{40}{\usefont{T1}{pag}{m}{n}{\small{P}}}~{\usefont{T1}{pag}{m}{n}{\small{x}}}\symbol{41}\symbol{46}\\
    {\usefont{T1}{pag}{b}{n}{\small{Proof}}}\symbol{46}\\
    \hphantom{ }\hphantom{ }{\usefont{T1}{pag}{m}{n}{\small{{\itshape eauto}}}}\symbol{46}\\
    {\usefont{T1}{pag}{b}{n}{\small{Qed}}}\symbol{46}}}
  \end{displaymath}
  \par
    With a modality we can define the type of modal propositions, where the {\usefont{T1}{pag}{m}{n}{\small{unit}}} law is actually an equivalence (modalities are closure operators, by the {\usefont{T1}{pag}{m}{n}{\small{join}}} law, so the type of modal propositions is the image of {\usefont{T1}{pag}{m}{n}{\small{M}}} up to logical equivalence).
    \begin{displaymath}
    \makebox[1.\linewidth]{\makebox[0.1\linewidth]{}\parbox{0.9\linewidth}{{\usefont{T1}{pag}{b}{n}{\small{Definition}}}~{\usefont{T1}{pag}{m}{n}{\small{MProp}}}~$\coloneqq $~\{~{\usefont{T1}{pag}{m}{n}{\small{P}}}\symbol{58}{\usefont{T1}{pag}{b}{n}{\small{Prop}}}~\hspace{0.1em}\rule[-0.6ex]{1.sp}{1.\baselineskip}~{\usefont{T1}{pag}{m}{n}{\small{M}}}~{\usefont{T1}{pag}{m}{n}{\small{P}}}~${\rightarrow}$~{\usefont{T1}{pag}{m}{n}{\small{P}}}~\}\symbol{46}}}
  \end{displaymath}
  \par
    Despite not being a sort, {\usefont{T1}{pag}{m}{n}{\small{MProp}}} can be seen as a subtype of {\usefont{T1}{pag}{b}{n}{\small{Prop}}} and, therefore, as a universe in the sense of Section~\ref{latex_lib_label_2}.
    \begin{displaymath}
    \makebox[1.\linewidth]{\makebox[0.1\linewidth]{}\parbox{0.9\linewidth}{{\usefont{T1}{pag}{b}{n}{\small{Definition}}}~{\usefont{T1}{pag}{m}{n}{\small{El}}}~\symbol{40}{\usefont{T1}{pag}{m}{n}{\small{P}}}\symbol{58}{\usefont{T1}{pag}{m}{n}{\small{MProp}}}\symbol{41}~\symbol{58}~{\usefont{T1}{pag}{b}{n}{\small{Prop}}}~$\coloneqq $~{\usefont{T1}{pag}{m}{n}{\small{proj1\_sig}}}~{\usefont{T1}{pag}{m}{n}{\small{P}}}\symbol{46}}}
  \end{displaymath}
  \par
  Because of {\usefont{T1}{pag}{m}{n}{\small{strength}}}, the {\usefont{T1}{pag}{m}{n}{\small{MProp}}} universe is closed by products of arbitrary types. The {\usefont{T1}{pag}{b}{n}{\small{Program}}} keyword makes it possible to populate {\usefont{T1}{pag}{m}{n}{\small{MProp}}} by giving the proposition {\usefont{T1}{pag}{m}{n}{\small{P}}} (\emph{first projection}) explicitly and discharging the proof that {\usefont{T1}{pag}{m}{n}{\small{M}}}~{\usefont{T1}{pag}{m}{n}{\small{P}}}~${\rightarrow}$~{\usefont{T1}{pag}{m}{n}{\small{P}}} to tactics.
  \begin{displaymath}
    \makebox[1.\linewidth]{\makebox[0.1\linewidth]{}\parbox{0.9\linewidth}{{\usefont{T1}{pag}{b}{n}{\small{Program}}}~{\usefont{T1}{pag}{b}{n}{\small{Definition}}}~{\usefont{T1}{pag}{m}{n}{\small{Forall}}}~\symbol{40}{\usefont{T1}{pag}{m}{n}{\small{A}}}\symbol{58}{\usefont{T1}{pag}{b}{n}{\small{Type}}}\symbol{41}~\symbol{40}{\usefont{T1}{pag}{m}{n}{\small{F}}}\symbol{58}{\usefont{T1}{pag}{m}{n}{\small{A}}}${\rightarrow}${\usefont{T1}{pag}{m}{n}{\small{MProp}}}\symbol{41}~\symbol{58}~{\usefont{T1}{pag}{m}{n}{\small{MProp}}}~$\coloneqq $\\
    \hphantom{ }\hphantom{ }{\usefont{T1}{pag}{b}{n}{\small{forall}}}~{\usefont{T1}{pag}{m}{n}{\small{x}}}\symbol{58}{\usefont{T1}{pag}{m}{n}{\small{A}}}\symbol{44}~{\usefont{T1}{pag}{m}{n}{\small{El}}}~\symbol{40}{\usefont{T1}{pag}{m}{n}{\small{F}}}~{\usefont{T1}{pag}{m}{n}{\small{x}}}\symbol{41}\symbol{46}\\
    {\usefont{T1}{pag}{b}{n}{\small{Next}}}~{\usefont{T1}{pag}{b}{n}{\small{Obligation}}}\symbol{46}\\
    \hphantom{ }\hphantom{ }{\usefont{T1}{pag}{m}{n}{\small{{\itshape intros}}}}~{\usefont{T1}{pag}{m}{n}{\small{A}}}~{\usefont{T1}{pag}{m}{n}{\small{F}}}~{\usefont{T1}{pag}{m}{n}{\small{h}}}~{\usefont{T1}{pag}{m}{n}{\small{x}}}\symbol{46}\\
    \hphantom{ }\hphantom{ }{\usefont{T1}{pag}{m}{n}{\small{{\itshape apply}}}}~{\usefont{T1}{pag}{m}{n}{\small{strength}}}~{\usefont{T1}{pag}{b}{n}{\small{with}}}~\symbol{40}{\usefont{T1}{pag}{m}{n}{\small{x}}}$\coloneqq ${\usefont{T1}{pag}{m}{n}{\small{x}}}\symbol{41}~{\usefont{T1}{pag}{b}{n}{\small{in}}}~{\usefont{T1}{pag}{m}{n}{\small{h}}}\symbol{46}\\
    \hphantom{ }\hphantom{ }{\usefont{T1}{pag}{m}{n}{\small{{\itshape destruct}}}}~\symbol{40}{\usefont{T1}{pag}{m}{n}{\small{F}}}~{\usefont{T1}{pag}{m}{n}{\small{x}}}\symbol{41}\symbol{46}~{\usefont{T1}{pag}{m}{n}{\small{cbn}}}~{\usefont{T1}{pag}{b}{n}{\small{in}}}~$*$\symbol{46}\\
    \hphantom{ }\hphantom{ }{\usefont{T1}{pag}{m}{n}{\small{{\itshape eauto}}}}\symbol{46}\\
    {\usefont{T1}{pag}{b}{n}{\small{Qed}}}\symbol{46}}}
  \end{displaymath}
  \par
    Definitions of $\mbox{\textrm{U}}^-$ products, small and large, for {\usefont{T1}{pag}{m}{n}{\small{MProp}}} follow immediately:
    \begin{displaymath}
    \makebox[1.\linewidth]{\makebox[0.1\linewidth]{}\parbox{0.9\linewidth}{{\usefont{T1}{pag}{b}{n}{\small{Let}}}~{\usefont{T1}{pag}{m}{n}{\small{Forall1}}}~\symbol{40}{\usefont{T1}{pag}{m}{n}{\small{u}}}\symbol{58}{\usefont{T1}{pag}{m}{n}{\small{MProp}}}\symbol{41}~\symbol{40}{\usefont{T1}{pag}{m}{n}{\small{F}}}\symbol{58}{\usefont{T1}{pag}{m}{n}{\small{El}}}~{\usefont{T1}{pag}{m}{n}{\small{u}}}~${\rightarrow}$~{\usefont{T1}{pag}{m}{n}{\small{MProp}}}\symbol{41}~\symbol{58}~{\usefont{T1}{pag}{m}{n}{\small{MProp}}}~$\coloneqq $~{\usefont{T1}{pag}{m}{n}{\small{Forall}}}~\symbol{40}{\usefont{T1}{pag}{m}{n}{\small{El}}}~{\usefont{T1}{pag}{m}{n}{\small{u}}}\symbol{41}~{\usefont{T1}{pag}{m}{n}{\small{F}}}\symbol{46}\\
    {\usefont{T1}{pag}{b}{n}{\small{Let}}}~{\usefont{T1}{pag}{m}{n}{\small{ForallU1}}}~\symbol{40}{\usefont{T1}{pag}{m}{n}{\small{F}}}\symbol{58}{\usefont{T1}{pag}{m}{n}{\small{MProp}}}${\rightarrow}${\usefont{T1}{pag}{m}{n}{\small{MProp}}}\symbol{41}~\symbol{58}~{\usefont{T1}{pag}{m}{n}{\small{MProp}}}~$\coloneqq $~{\usefont{T1}{pag}{m}{n}{\small{Forall}}}~{\usefont{T1}{pag}{m}{n}{\small{MProp}}}~{\usefont{T1}{pag}{m}{n}{\small{F}}}\symbol{46}}}
  \end{displaymath}
    Because $\mbox{{\usefont{T1}{pag}{m}{n}{\small{El}}}~\symbol{40}{\usefont{T1}{pag}{m}{n}{\small{Forall}}}~{\usefont{T1}{pag}{m}{n}{\small{A}}}~{\usefont{T1}{pag}{m}{n}{\small{F}}}\symbol{41}}=\mbox{{\usefont{T1}{pag}{b}{n}{\small{forall}}}~{\usefont{T1}{pag}{m}{n}{\small{x}}}\symbol{58}{\usefont{T1}{pag}{m}{n}{\small{A}}}\symbol{44}~{\usefont{T1}{pag}{m}{n}{\small{F}}}}$, introduction, elimination and ${\beta}$-rules for the products are immediate.
  \par
    Just like in Section~\ref{latex_lib_label_5}, a retraction of {\usefont{T1}{pag}{m}{n}{\small{MProp}}} into a modal proposition can be used to trigger Hurkens's paradox. This is an example of instance of Hurkens's paradox where neither of the universes are sorts of the system.
  \par
    \begin{displaymath}
    \makebox[1.\linewidth]{\makebox[0.1\linewidth]{}\parbox{0.9\linewidth}{{\usefont{T1}{pag}{b}{n}{\small{Variable}}}~{\usefont{T1}{pag}{m}{n}{\small{U0}}}\symbol{58}{\usefont{T1}{pag}{m}{n}{\small{MProp}}}\symbol{46}\\
    {\usefont{T1}{pag}{b}{n}{\small{Variable}}}~{\usefont{T1}{pag}{m}{n}{\small{proj0}}}~\symbol{58}~{\usefont{T1}{pag}{m}{n}{\small{U0}}}~${\rightarrow}$~{\usefont{T1}{pag}{m}{n}{\small{MProp}}}\symbol{46}\\
    {\usefont{T1}{pag}{b}{n}{\small{Variable}}}~{\usefont{T1}{pag}{m}{n}{\small{inj0}}}~\symbol{58}~{\usefont{T1}{pag}{m}{n}{\small{MProp}}}~${\rightarrow}$~{\usefont{T1}{pag}{m}{n}{\small{U0}}}\symbol{46}\\
    {\usefont{T1}{pag}{b}{n}{\small{Hypothesis}}}~{\usefont{T1}{pag}{m}{n}{\small{inj0\_unit}}}~\symbol{58}~{\usefont{T1}{pag}{b}{n}{\small{forall}}}~\symbol{40}{\usefont{T1}{pag}{m}{n}{\small{A}}}\symbol{58}{\usefont{T1}{pag}{m}{n}{\small{MProp}}}\symbol{41}\symbol{44}~{\usefont{T1}{pag}{m}{n}{\small{El}}}~{\usefont{T1}{pag}{m}{n}{\small{A}}}~${\rightarrow}$~{\usefont{T1}{pag}{m}{n}{\small{El}}}~\symbol{40}{\usefont{T1}{pag}{m}{n}{\small{proj0}}}~\symbol{40}{\usefont{T1}{pag}{m}{n}{\small{inj0}}}~{\usefont{T1}{pag}{m}{n}{\small{A}}}\symbol{41}\symbol{41}\symbol{46}\\
    {\usefont{T1}{pag}{b}{n}{\small{Hypothesis}}}~{\usefont{T1}{pag}{m}{n}{\small{inj0\_counit}}}~\symbol{58}~{\usefont{T1}{pag}{b}{n}{\small{forall}}}~\symbol{40}{\usefont{T1}{pag}{m}{n}{\small{A}}}\symbol{58}{\usefont{T1}{pag}{m}{n}{\small{MProp}}}\symbol{41}\symbol{44}~{\usefont{T1}{pag}{m}{n}{\small{El}}}~\symbol{40}{\usefont{T1}{pag}{m}{n}{\small{proj0}}}~\symbol{40}{\usefont{T1}{pag}{m}{n}{\small{inj0}}}~{\usefont{T1}{pag}{m}{n}{\small{A}}}\symbol{41}\symbol{41}~${\rightarrow}$~{\usefont{T1}{pag}{m}{n}{\small{El}}}~{\usefont{T1}{pag}{m}{n}{\small{A}}}\symbol{46}}}
  \end{displaymath}
  \par
    Following the the proof of Section~\ref{latex_lib_label_5}, we conclude from this context that every modal proposition is inhabited. This is not necessarily a contradiction, as falsity need not be modal. For instance the trivial modality, whose only modal proposition in {\usefont{T1}{pag}{m}{n}{\small{True}}}.
    \begin{displaymath}
    \makebox[1.\linewidth]{\makebox[0.1\linewidth]{}\parbox{0.9\linewidth}{{\usefont{T1}{pag}{b}{n}{\small{Definition}}}~{\usefont{T1}{pag}{m}{n}{\small{M}}}~\symbol{40}{\usefont{T1}{pag}{m}{n}{\small{A}}}\symbol{58}{\usefont{T1}{pag}{b}{n}{\small{Prop}}}\symbol{41}~\symbol{58}~{\usefont{T1}{pag}{b}{n}{\small{Prop}}}~$\coloneqq $~{\usefont{T1}{pag}{m}{n}{\small{True}}}}}
  \end{displaymath}
  \par
    A more interesting modality is, for a given {\usefont{T1}{pag}{m}{n}{\small{X}}}:
    \begin{displaymath}
    \makebox[1.\linewidth]{\makebox[0.1\linewidth]{}\parbox{0.9\linewidth}{{\usefont{T1}{pag}{b}{n}{\small{Definition}}}~{\usefont{T1}{pag}{m}{n}{\small{M}}}~\symbol{40}{\usefont{T1}{pag}{m}{n}{\small{A}}}\symbol{58}{\usefont{T1}{pag}{b}{n}{\small{Prop}}}\symbol{41}~\symbol{58}~{\usefont{T1}{pag}{b}{n}{\small{Prop}}}~$\coloneqq $~{\usefont{T1}{pag}{m}{n}{\small{A}}}${\lor}${\usefont{T1}{pag}{m}{n}{\small{X}}}}}
  \end{displaymath}
    for such a modality exhibiting a retraction into a modal proposition only prove ${\lnot}${\usefont{T1}{pag}{m}{n}{\small{X}}}: it is always the case that the smallest modal proposition is {\usefont{T1}{pag}{m}{n}{\small{M}}}~{\usefont{T1}{pag}{m}{n}{\small{False}}}.
  \par
    \section{Weak excluded middle and proof irrelevance}
  \par
    In this section we will be concerned with the double-negation modality, whose modal propositions are also called \emph{negative propositions}:
    \begin{displaymath}
    \makebox[1.\linewidth]{\makebox[0.1\linewidth]{}\parbox{0.9\linewidth}{{\usefont{T1}{pag}{b}{n}{\small{Definition}}}~{\usefont{T1}{pag}{m}{n}{\small{M}}}~\symbol{40}{\usefont{T1}{pag}{m}{n}{\small{A}}}\symbol{58}{\usefont{T1}{pag}{b}{n}{\small{Prop}}}\symbol{41}~\symbol{58}~{\usefont{T1}{pag}{b}{n}{\small{Prop}}}~$\coloneqq $~${\lnot}$${\lnot}${\usefont{T1}{pag}{m}{n}{\small{A}}}}}
  \end{displaymath}
    and will use the paradox from Section~\ref{latex_lib_label_7}, to prove that the weak principle of excluded middle
    \begin{displaymath}
    \makebox[1.\linewidth]{\makebox[0.1\linewidth]{}\parbox{0.9\linewidth}{{\usefont{T1}{pag}{b}{n}{\small{Hypothesis}}}~{\usefont{T1}{pag}{m}{n}{\small{wem}}}~\symbol{58}~{\usefont{T1}{pag}{b}{n}{\small{forall}}}~{\usefont{T1}{pag}{m}{n}{\small{A}}}\symbol{58}{\usefont{T1}{pag}{b}{n}{\small{Prop}}}\symbol{44}~${\lnot}$${\lnot}${\usefont{T1}{pag}{m}{n}{\small{A}}}~${\lor}$~${\lnot}${\usefont{T1}{pag}{m}{n}{\small{A}}}\symbol{46}}}
  \end{displaymath}
    entails a weak form of proof irrelevance. This is a new proof I added to \textsf{theories/Logic/ClassicalFact.v} and is available from version 8.5.
  \par
    Looking closely at {\usefont{T1}{pag}{m}{n}{\small{wem}}} it becomes clear that it claims decidability of exactly the negative propositions.
    \begin{displaymath}
    \makebox[1.\linewidth]{\makebox[0.1\linewidth]{}\parbox{0.9\linewidth}{{\usefont{T1}{pag}{b}{n}{\small{Remark}}}~{\usefont{T1}{pag}{m}{n}{\small{wem}}}\symbol{39}~\symbol{58}~{\usefont{T1}{pag}{b}{n}{\small{forall}}}~{\usefont{T1}{pag}{m}{n}{\small{A}}}\symbol{58}{\usefont{T1}{pag}{m}{n}{\small{MProp}}}\symbol{44}~{\usefont{T1}{pag}{m}{n}{\small{El}}}~{\usefont{T1}{pag}{m}{n}{\small{A}}}~${\lor}$~${\lnot}${\usefont{T1}{pag}{m}{n}{\small{El}}}~{\usefont{T1}{pag}{m}{n}{\small{A}}}\symbol{46}}}
  \end{displaymath}
  \par
    The proof, therefore, proceeds just like the  proof of Section~\ref{latex_lib_label_6}. We begin by postulating a proposition with two proofs.
    \begin{displaymath}
    \makebox[1.\linewidth]{\makebox[0.1\linewidth]{}\parbox{0.9\linewidth}{{\usefont{T1}{pag}{b}{n}{\small{Variable}}}~{\usefont{T1}{pag}{m}{n}{\small{U0}}}\symbol{58}{\usefont{T1}{pag}{b}{n}{\small{Prop}}}\symbol{46}\\
    {\usefont{T1}{pag}{b}{n}{\small{Variables}}}~{\usefont{T1}{pag}{m}{n}{\small{t}}}~{\usefont{T1}{pag}{m}{n}{\small{f}}}~\symbol{58}~{\usefont{T1}{pag}{m}{n}{\small{U0}}}\symbol{46}\\
    {\usefont{T1}{pag}{b}{n}{\small{Hypothesis}}}~{\usefont{T1}{pag}{m}{n}{\small{not\_eq\_t\_f}}}~\symbol{58}~{\usefont{T1}{pag}{m}{n}{\small{t}}}~${\ne}$~{\usefont{T1}{pag}{m}{n}{\small{f}}}\symbol{46}}}
  \end{displaymath}
  \par
    Notice that {\usefont{T1}{pag}{m}{n}{\small{U0}}} is negative, since {\usefont{T1}{pag}{m}{n}{\small{U0}}} has a proof, in particular ${\lnot}$${\lnot}${\usefont{T1}{pag}{m}{n}{\small{U0}}}${\rightarrow}${\usefont{T1}{pag}{m}{n}{\small{U0}}} holds. So we only need to construct a retraction into {\usefont{T1}{pag}{m}{n}{\small{U0}}}. The retraction is given by {\usefont{T1}{pag}{m}{n}{\small{inj0}}} and {\usefont{T1}{pag}{m}{n}{\small{proj0}}} which are, \emph{mutatis mutandis} the same as in Section~\ref{latex_lib_label_6}: double negations have to be inserted for propositions which need to be negative, and proofs of negativity have to be provided when building negative propositions.
    \begin{displaymath}
    \makebox[1.\linewidth]{\makebox[0.1\linewidth]{}\parbox{0.9\linewidth}{{\usefont{T1}{pag}{b}{n}{\small{Let}}}~{\usefont{T1}{pag}{m}{n}{\small{inj0}}}~\symbol{40}{\usefont{T1}{pag}{m}{n}{\small{A}}}\symbol{58}{\usefont{T1}{pag}{m}{n}{\small{MProp}}}\symbol{41}~\symbol{58}~{\usefont{T1}{pag}{m}{n}{\small{U0}}}~$\coloneqq $\\
    \hphantom{ }\hphantom{ }\hphantom{ }\hphantom{ }\hphantom{ }{\usefont{T1}{pag}{m}{n}{\small{or\_ind}}}~\symbol{40}{\usefont{T1}{pag}{m}{n}{\small{El}}}~{\usefont{T1}{pag}{m}{n}{\small{A}}}\symbol{41}~\symbol{40}${\lnot}${\usefont{T1}{pag}{m}{n}{\small{El}}}~{\usefont{T1}{pag}{m}{n}{\small{A}}}\symbol{41}~{\usefont{T1}{pag}{m}{n}{\small{U0}}}~\symbol{40}{\usefont{T1}{pag}{b}{n}{\small{fun}}}~$\_$~${\Rightarrow}$~{\usefont{T1}{pag}{m}{n}{\small{t}}}\symbol{41}~\symbol{40}{\usefont{T1}{pag}{b}{n}{\small{fun}}}~$\_$~${\Rightarrow}$~{\usefont{T1}{pag}{m}{n}{\small{f}}}\symbol{41}~\symbol{40}{\usefont{T1}{pag}{m}{n}{\small{wem}}}\symbol{39}~{\usefont{T1}{pag}{m}{n}{\small{A}}}\symbol{41}\symbol{46}\\
    {\usefont{T1}{pag}{b}{n}{\small{Let}}}~{\usefont{T1}{pag}{m}{n}{\small{proj0}}}~\symbol{40}{\usefont{T1}{pag}{m}{n}{\small{x}}}\symbol{58}{\usefont{T1}{pag}{m}{n}{\small{U0}}}\symbol{41}~\symbol{58}~{\usefont{T1}{pag}{m}{n}{\small{MProp}}}~$\coloneqq $\\
    \hphantom{ }\hphantom{ }\hphantom{ }\hphantom{ }\hphantom{ }{\usefont{T1}{pag}{m}{n}{\small{exist}}}~\symbol{40}{\usefont{T1}{pag}{b}{n}{\small{fun}}}~{\usefont{T1}{pag}{m}{n}{\small{P}}}${\Rightarrow}$${\lnot}$${\lnot}${\usefont{T1}{pag}{m}{n}{\small{P}}}~${\rightarrow}$~{\usefont{T1}{pag}{m}{n}{\small{P}}}\symbol{41}~\symbol{40}${\lnot}$${\lnot}$\symbol{40}{\usefont{T1}{pag}{m}{n}{\small{t}}}~\symbol{61}~{\usefont{T1}{pag}{m}{n}{\small{x}}}\symbol{41}\symbol{41}~\symbol{40}{\usefont{T1}{pag}{b}{n}{\small{fun}}}~{\usefont{T1}{pag}{m}{n}{\small{h}}}~{\usefont{T1}{pag}{m}{n}{\small{x}}}~${\Rightarrow}$~{\usefont{T1}{pag}{m}{n}{\small{h}}}~\symbol{40}{\usefont{T1}{pag}{b}{n}{\small{fun}}}~{\usefont{T1}{pag}{m}{n}{\small{k}}}~${\Rightarrow}$~{\usefont{T1}{pag}{m}{n}{\small{k}}}~{\usefont{T1}{pag}{m}{n}{\small{x}}}\symbol{41}\symbol{41}\symbol{46}}}
  \end{displaymath}
  \par
    The unit and counit laws follow and we eventually derive a contradiction. That is, since {\usefont{T1}{pag}{m}{n}{\small{U0}}}\symbol{58}{\usefont{T1}{pag}{b}{n}{\small{Prop}}} is arbitrary a proof that:
    \begin{displaymath}
    \makebox[1.\linewidth]{\makebox[0.1\linewidth]{}\parbox{0.9\linewidth}{{\usefont{T1}{pag}{b}{n}{\small{forall}}}~\symbol{40}{\usefont{T1}{pag}{m}{n}{\small{A}}}\symbol{58}{\usefont{T1}{pag}{b}{n}{\small{Prop}}}\symbol{41}~\symbol{40}{\usefont{T1}{pag}{m}{n}{\small{x}}}~{\usefont{T1}{pag}{m}{n}{\small{y}}}\symbol{58}{\usefont{T1}{pag}{m}{n}{\small{A}}}\symbol{41}\symbol{44}~${\lnot}$${\lnot}$\symbol{40}{\usefont{T1}{pag}{m}{n}{\small{x}}}\symbol{61}{\usefont{T1}{pag}{m}{n}{\small{y}}}\symbol{41}}}
  \end{displaymath}
  \par
    Contrary to to the case of (strong) excluded middle, we cannot eliminate this last double-negative. So proof irrelevance doesn't follow from weak excluded middle. However, this section proves that weak excluded middle is incompatible with any sort of proof relevance principle. In particular, in Coq lingo, weak excluded middle cannot hold in impredicative {\usefont{T1}{pag}{m}{n}{\small{Set}}}, that is an impredicative sort with strong elimination.
  \chapter{Conclusion}
  \par
    The axiomatisation of Hurkens's paradox presented in Section~\ref{latex_lib_label_1} is very versatile. It can be used, mostly, to prove that some combination of logical principles are incompatible, but also to detect bugs in a dependent-type-theory implementation. Which is a completely fair and healthy activity if you ask this author.
  \par
    It is, certainly, an improvement over a situation where each paradox would need a careful redesign of Hurkens's proof to fit the specific premises. In practice it meant that paradoxes were not derived, because the brave paradox-finder didn't have the energy or expertise to translate Hurkens's paradox.
  \par
  As per the axiomatisation itself. It has the pleasant property of requiring only a subset of $\mbox{\textrm{U}}^-$ where the ``proofs'' of ``propositions'' don't require ${\beta}$-rules or any kind of equality rule. So something was learned. Adapting the proof to the axiomatisation doesn't present any new difficulty, except from controlling rewriting a little. It wasn't discovered before solely by the virtue of nobody looking. The reader who enjoyed this axiomatisation can celebrate the bout of optimism which made me look the right way, and the night I lost over it.
    \bibliography{library}

\begin{thebibliography}{1}

\bibitem{Barendregt1991}
Henk Barendregt.
\newblock {Introduction to generalized type systems}.
\newblock {\em Journal of Functional Programming}, 1(2):125--154, 1991.

\bibitem{Barendregt}
Henk Barendregt.
\newblock {Lambda calculi with types}.
\newblock {\em Handbook of logic in computer science}, 1992.

\bibitem{Coquand86}
Thierry Coquand.
\newblock {An analysis of Girard's paradox}.
\newblock Technical report, 1986.

\bibitem{Coquand1989}
Thierry Coquand.
\newblock {Metamathematical investigations of a calculus of constructions}.
\newblock Technical report, INRIA, 1989.

\bibitem{Geuvers2007}
Herman Geuvers.
\newblock {Inconsistency of classical logic in type theory}.
\newblock 2007.

\bibitem{Girard1972}
Jean-Yves Girard.
\newblock {\em {Interpr\'{e}tation fonctionnelle et \'{e}limination des
  coupures de l'arithm\'{e}tique d'ordre sup\'{e}rieur}}.
\newblock Th\`{e}se d'\'{E}tat, Paris 7, 1972.

\bibitem{HerbelinSpiwack2013}
Hugo Herbelin and Arnaud Spiwack.
\newblock {The Rooster and the Syntactic Bracket}.
\newblock In Ralph Matthes and Aleksy Schubert, editors, {\em 19th
  International Conference on Types for Proofs and Programs (TYPES 2013)},
  volume~26 of {\em Leibniz International Proceedings in Informatics (LIPIcs)},
  pages 169----187, Dagstuhl, Germany, 2014. Schloss Dagstuhl--Leibniz-Zentrum
  fuer Informatik.

\bibitem{Hurkens1995}
Antonius Hurkens.
\newblock {A simplification of Girard's paradox}.
\newblock {\em Typed Lambda Calculi and Applications}, pages 266--278, 1995.

\bibitem{Coq}
{The Coq development team}.
\newblock {The Coq Proof Assistant}.
\newblock http://coq.inria.fr/.

\end{thebibliography}
\end{document}